\documentclass[11pt,letterpaper]{article}
\usepackage[T1]{fontenc}
\usepackage{lmodern}
\usepackage[margin=0.85in]{geometry}
\usepackage{amsmath,amssymb}
\usepackage{graphicx}
\usepackage{array,booktabs,float}
\usepackage[font=small,labelfont=bf,labelsep=period]{caption}
\usepackage{fancyvrb}
\usepackage[normalem]{ulem}
\usepackage[super,sort&compress]{natbib}
\setcitestyle{citesep={,}}
\usepackage{xurl}
\usepackage[hidelinks]{hyperref}
\usepackage[nopatch=footnote]{microtype}
\providecommand{\ul}[1]{\uline{#1}}
\DeclareUnicodeCharacter{FF1A}{\makebox[1em]{:}}
\DeclareUnicodeCharacter{FF08}{(}
\DeclareUnicodeCharacter{FF09}{)}

\begin{document}

\begin{center}
{\LARGE\bfseries \textbf{CGTransformer: A Topology-Aware Autoregressive Framework for Crystal Generation and Inverse Design}\par}
\medskip
Zhennan Zhang\textsuperscript{1, \textdagger{}}, Shaobo Yu\textsuperscript{1, \textdagger{}}, Haoting Zhang\textsuperscript{1, \textdagger{}}, Yu Han\textsuperscript{1}, Zhiyue Guo\textsuperscript{1}, Junjie Wang\textsuperscript{1,*}, Hao Gao\textsuperscript{2,3,*}, Jian Sun\textsuperscript{1,*}\par
\smallskip
\small
\textsuperscript{1}~National Laboratory of Solid State Microstructures, School of Physics and Collaborative Innovation Center of Advanced Microstructures, Nanjing University, Nanjing 210093, China\par
\textsuperscript{2}~Guangdong Basic Research Center of Excellence for Structure and Fundamental Interactions of Matter, Guangdong Provincial Key Laboratory of Quantum Engineering and Quantum Materials, School of Physics, South China Normal University, Guangzhou 510006, China\par
\textsuperscript{3}~Guangdong-Hong Kong Joint Laboratory of Quantum Matter, Frontier Research Institute for Physics, South China Normal University, Guangzhou 510006, China\par
\end{center}
\begingroup
\renewcommand{\thefootnote}{}
\footnotetext[0]{\textsuperscript{\textdagger{}} These authors contributed equally to this work.

\textsuperscript{*}Corresponding author. J.S (\href{mailto:jiansun@nju.edu.cn}{\nolinkurl{jiansun@nju.edu.cn}}); H.G. (\href{mailto:gaohao@m.scnu.edu.cn}{\nolinkurl{gaohao@m.scnu.edu.cn}}); J.W. (\href{mailto:wangjunjie@nju.edu.cn}{\nolinkurl{wangjunjie@nju.edu.cn}})}
\endgroup

\begin{abstract}
\raggedright
Autoregressive language models offer a unified route to crystal generation and inverse design, but their performance depends on how periodic structures are serialized. Symmetry-equivalent and closely related crystals can receive different atomic orders under coordinate-based serialization, introducing variability unrelated to their structural connectivity. Here, we present CGTransformer (Crystal Graph Transformer), an autoregressive generative framework that addresses this challenge via a graph-embedded sorting strategy. Using quotient-graph topology and barycentric embeddings, Element--Symmetry--Graph (ESG) converts periodic crystal networks into topology-guided one-dimensional sequences that retain key local and symmetry-related structural information. CGTransformer achieves high generation validity and the highest single-shot crystal structure prediction (CSP) match rate among the compared MP-20 baselines (63.24\%), while same-backbone comparisons show improved structure recovery over coordinate-based ordering. Furthermore, we demonstrate property-conditioned inverse design for wide-bandgap semiconductors, where ESG achieves a much higher stable, unique and novel (S.U.N.) rate and stronger target-property alignment than the coordinate-based baseline. These results establish topology-guided sequence construction as an effective route to improving crystal generation and property-directed design at fixed model capacity.
\end{abstract}

\section{Introduction}

Discovering new crystalline materials is central to applications ranging from energy storage to catalysis. Traditional materials discovery largely follows a forward-screening paradigm, in which candidate structures are generated from known templates and evaluated using high-throughput density functional theory (DFT) calculations combined with evolutionary search algorithms\citep{ref01}. Although crystal structure prediction (CSP) methods\citep{ref02,ref03,ref04,ref05,ref06,ref07,ref08,ref09,ref10,ref11} have achieved notable success, their trial-and-error exploration of vast chemical and configurational spaces remains computationally demanding and often inefficient. Deep generative models represent the state-of-the-art approach for materials inverse design\citep{ref01,ref12,ref13}. At present, the landscape of generative materials design is highly diverse, encompassing several architectures, including variational autoencoders (VAE)\citep{ref14,ref15,ref16,ref17,ref18}, diffusion models\citep{ref19,ref20,ref21,ref22,ref23,ref24,ref25}, flow matching\citep{ref26,ref27}, large language models (LLM)\citep{ref12,ref28,ref29,ref30,ref31,ref32,ref33,ref34}.

Among these paradigms, LLMs are particularly attractive because they provide a unified autoregressive framework for both unconditional generation and conditional design, while naturally benefiting from advances in large-scale sequence modelling\citep{ref32,ref35}. Their application to crystals, however, is not straightforward, because the representation is inherently ambiguous: autoregressive models are inherently sensitive to sequence order, whereas crystals are periodic structures that can admit many symmetry-equivalent serializations\citep{ref29}. More generally, recent work in autoregressive modelling and graph theory has shown that order sensitivity can become a bottleneck when the underlying object does not possess a unique natural serialization\citep{ref36,ref37,ref38,ref39}. Recently, progress has been made to bridge this gap. For instance, models based on crystallographic text formats (e.g. cif or xyz files), such as CrystaLLM\citep{ref28}, demonstrate that end-to-end crystal generation is feasible within an autoregressive framework. To alleviate permutation explosion, coordinate-and-symmetry-based sorting methods such as the "Element-Symmetry-Position" (ESP) paradigm are adopted in CrystalFormer\citep{ref30} and Mat2Seq\citep{ref29}, where the atomic order is determined by atomic number, symmetry multiplicity, and the lexicographical ordering of fractional coordinates (i.e., \(x \rightarrow y \rightarrow z\)).They have successfully compressed the permutation space to a linear scale. Mat2Seq also achieves SE(3) and periodic invariance at the sequence level through mathematical transformations.

The central question is how the atomic ordering presented to an autoregressive model affects its ability to recover crystal structures and generate materials under property constraints. We address this question by treating sequence construction as an explicit representation-level design variable. On the one hand, methods based on raw CIF text are still troubled by permutation explosion. They only use a limited set of structural snapshots even with data augment\citep{ref29}. On the other hand, although symmetry- and coordinate-based strategies substantially reduce this ambiguity, their final lexicographical ordering of real-space fractional coordinates can change dramatically under small structural perturbations and may not reflect local bonding connectivity. This residual coordinate-induced instability introduces sequence variability that is unrelated to the underlying crystal topology and can therefore impose a misaligned inductive bias on autoregressive learning.

A natural way to reduce coordinate sensitivity is to represent crystals through quotient graphs, which fold infinite periodic networks into finite graphs while preserving their underlying connectivity. Quotient-graph and related graph-theoretic methods have long been used in crystallography to analyze periodic structures\citep{ref40,ref41,ref42}, and more recent studies have applied graph-based ideas to generate complex carbon allotropes with specific coordination numbers (e.g., in the RG2 method)\citep{ref43}, and used for modular and crystal structure decomposition to accelerate structure searches such as MAGUS\citep{ref09,ref10,ref11,ref44,ref45} as well as for the refinement and pre-selection of random structures\citep{ref46}. These developments suggest that topological properties may offer a more stable basis for sequence construction than direct ordering in real space. Recent work on materials property prediction shows that language models do not automatically exploit numerical coordinate information, even as model and dataset sizes increase\citep{ref47}. For crystal generation, these findings motivate using periodic topology to organize atomic ordering while retaining fractional coordinates and lattice parameters in the model sequence.

Methods like SLICES\citep{ref48} used quotient graphs to construct invariant string representations and have led to the discovery of new materials. However, the model input representation in SLICES still takes the form of a token sequence: the atom and edge orders are not intrinsically fixed, and the explicit edge vectors written into the sequence are themselves not translationally invariant. As a result, equivalent structures can still admit multiple valid serializations, and the original framework relies mainly on randomization-based augmentation to mitigate this order sensitivity rather than resolving it through a uniquely stabilized sequence construction. We therefore treat sequence construction as a central design problem, and seek a topology-aware representation that reduces nuisance variability before autoregressive learning.

Thus, we incorporate quotient-graph information into autoregressive crystal serialization and introduce the CGTransformer (Crystal Graph Transformer) framework. Inspired by previous studies\citep{ref49,ref50,ref51,ref52}, we define the sequence using periodic quotient graphs together with barycentric embeddings instead of ordering atoms directly in real space, yielding an Element--Symmetry--Graph (ESG) serialization strategy. ESG uses periodic connectivity to guide the ordering of symmetry-equivalent atoms, replacing direct coordinate ranking with a deterministic graph-guided ordering under the specified graph-construction and starting-node protocol. In this way, ESG reduces coordinate-induced permutation ambiguity and produces a more consistent sequence space that better preserves local chemical environments for autoregressive modelling, enabling autoregressive models to learn three-dimensional structural regularities and structure--property relationships more effectively.

We evaluate this framework on structural-prototype extraction, public benchmark datasets including MP-20\citep{ref53}, MPTS-52\citep{ref54} and Alex-MP-20\citep{ref21}, covering unconditional generation, crystal structure prediction and property-conditioned inverse design. Detailed benchmark and calculation settings are provided in Methods section and Supplementary Note S1. Across unconditional generation and CSP tasks, CGTransformer delivers strong performance with relatively modest model sizes. In inverse design of wide-bandgap semiconductors, it generates more stable, unique and novel (S.U.N.) candidate materials than an ESP-based baseline. Together, these results suggest that topology-aware sequence representations can play a central role in autoregressive modelling of crystalline materials.

\section{Results}

\subsection{Overview of the CGTransformer Framework}

\begin{figure}[H]
\centering
\includegraphics[width=1\linewidth]{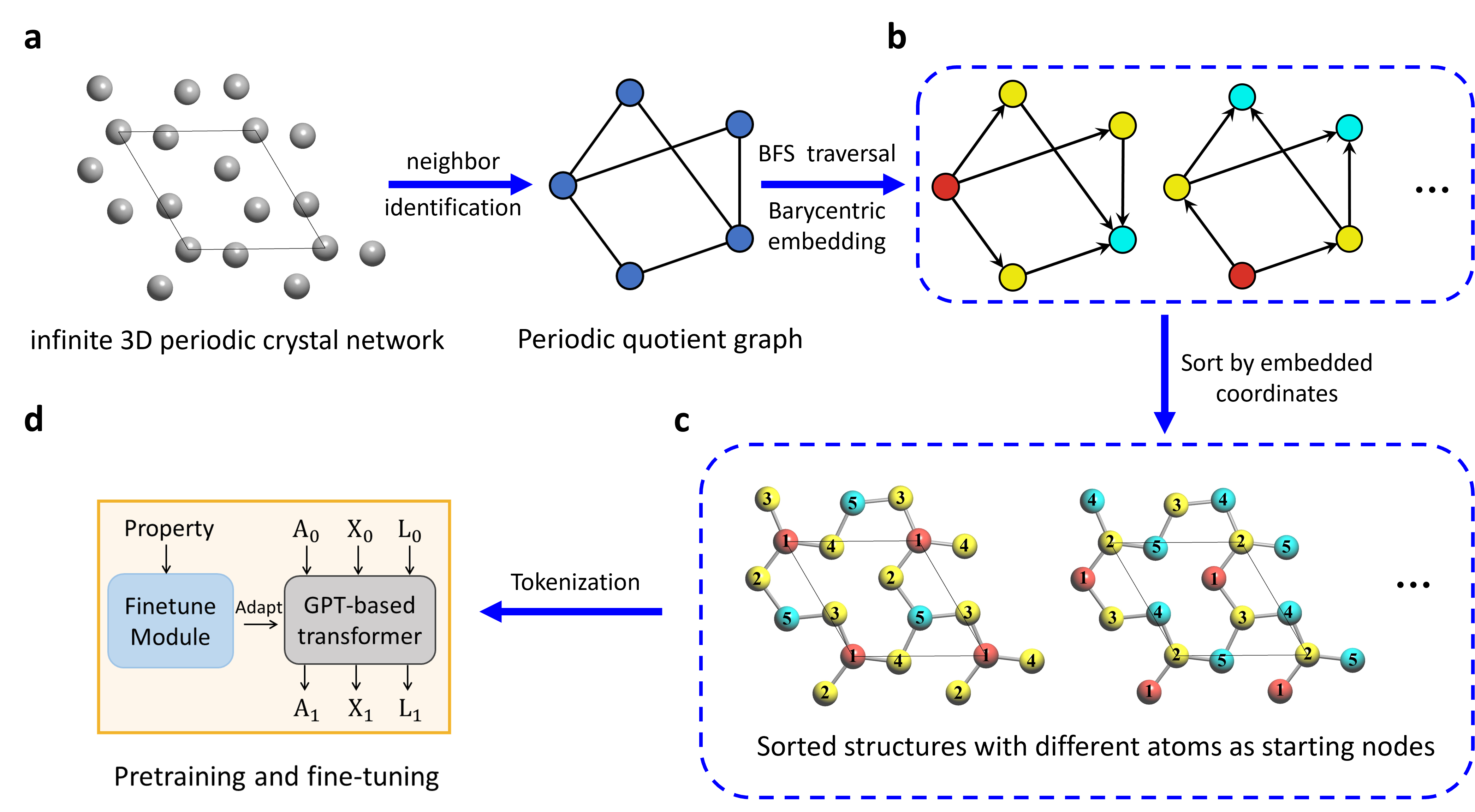}
\caption{\textbf{Overview of the CGTransformer framework and ESG-based crystal serialization.} \textbf{a,} A three-dimensional periodic crystal is converted into a finite periodic quotient graph by identifying local bonding environments. \textbf{b,} Breadth-first search (BFS) assigns topological depths to graph nodes; node colours indicate depth. \textbf{c,} For nodes with the same depth, ESG uses topology-derived coordinates from barycentric embedding to define a deterministic ordering that is less sensitive to small geometric perturbations than direct real-space sorting; numbers indicate the final atomic order. \textbf{d,} The resulting 1D representation, composed of element types, fractional coordinates and lattice vectors, is processed by a GPT-based autoregressive model. During fine-tuning, property tokens such as bandgap are prepended to enable property-conditioned crystal generation and inverse design.}
\label{fig:1}
\end{figure}

The overall generative and inverse design pipeline of the CGTransformer framework is illustrated in \textbf{Fig.~\ref{fig:1}}. First, the input 3D crystal network is systematically mapped into a finite periodic quotient graph via neighborhood identification. A breadth-first search (BFS) algorithm is then employed to determine the topological depth of each atom (indicated by the node colors in \textbf{Fig.~\ref{fig:1}}). For symmetry-equivalent nodes at the same depth, the ESG strategy introduces topology-derived coordinates for deterministic tie-breaking, thereby reducing the sensitivity of the ordering to small geometric perturbations. This procedure yields a topology-informed 1D sequence that better preserves local coordination relationships. Details of the sorting algorithm are provided in the Methods.

While the ESP and ESG strategies reduce the potential O(N!) permutations to at most O(N) topologically equivalent variants obtained by varying the starting atom, efficiently covering this intrinsic space with limited computational resources remains essential. To address this and prevent the model from overfitting to specific permutation orders, we adopt a dynamic stochastic sampling strategy inspired by GraphRNN\citep{ref55}. Specifically, in each training epoch, one valid sequence variant is randomly sampled for each crystal structure.

Finally, the physically consistent sequences comprising atom types (A), fractional coordinates (X), and lattice vectors (L) (see Methods and Supplementary Note S2 and Table S1 for details) are used to pretrain a GPT-2-based autoregressive transformer. For property-conditioned generation, physical attributes such as bandgap are converted into condition tokens and prepended to the input sequence during fine-tuning. This design enables both unconditional crystal generation and conditional inverse design within a unified autoregressive framework. Further details regarding the model architecture and training details are provided in Supplementary Note S3 and Supplementary Table S2.

\subsection{Structure Prototypes}

Before evaluating the generative performance of CGTransformer, we first examine whether the sequence representation remains stable across isostructural materials. In practice, many crystal families share the same structural prototype, that is, the same space-group symmetry and underlying bonding topology, while differing in chemical substitution or undergoing small local distortions. Such perturbations modify bond lengths and angles in real space but do not necessarily change the underlying connectivity.

\begin{figure}[H]
\centering
\includegraphics[width=1\linewidth]{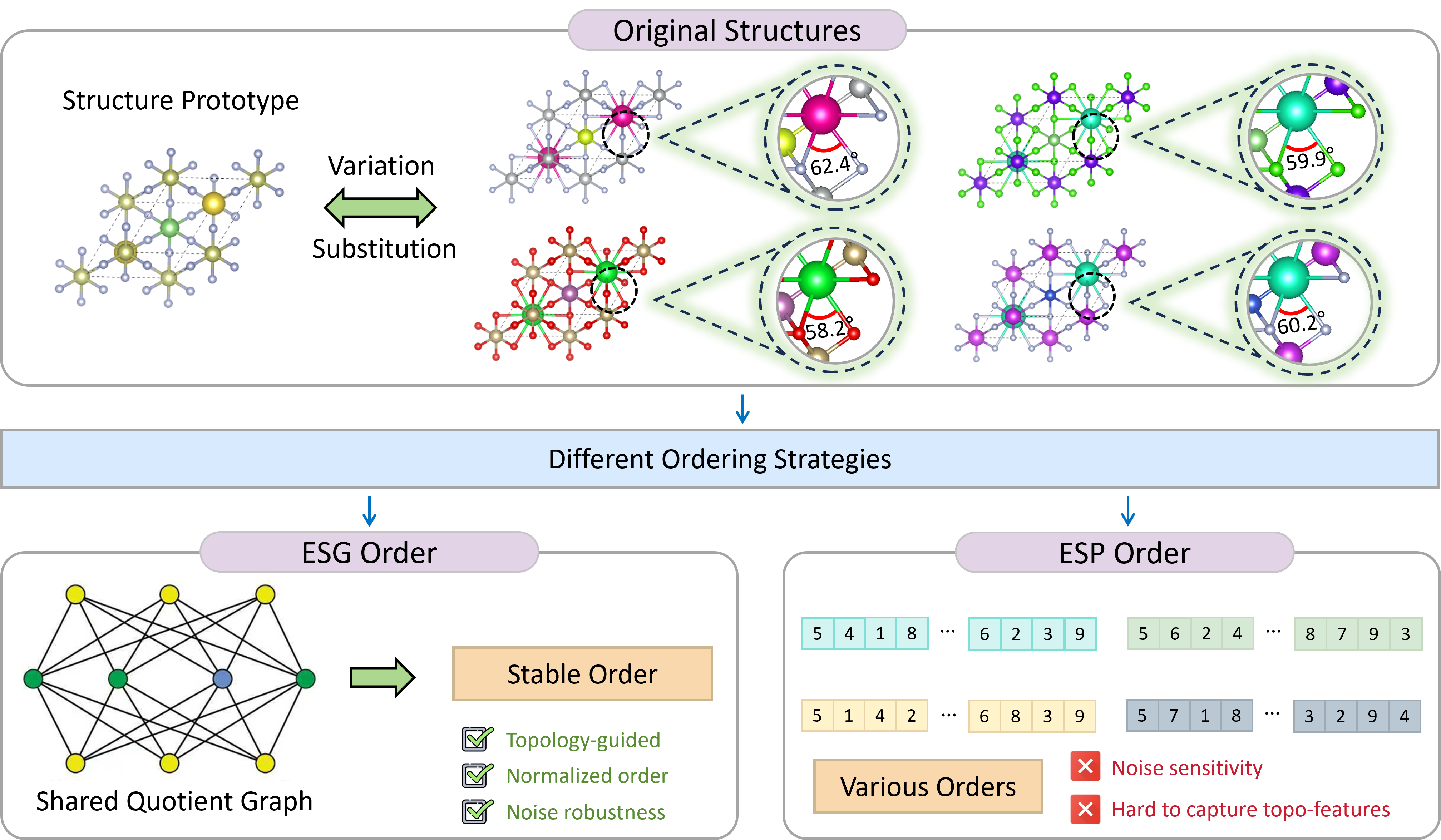}
\caption{\textbf{ESG reduces ordering variability across related crystal structures.} Top, representative structures derived from a common structural prototype through substitution or local relaxation retain similar underlying connectivity despite small changes in bond lengths and angles. Bottom left, ESG uses quotient-graph information to define a topology-aware ordering that is less sensitive to local geometric perturbations. Bottom right, ESP relies on direct real-space sorting, so small coordinate variations can produce different serialized orders for related structures.}
\label{fig:2}
\end{figure}

Because ESP depends on lexicographic sorting of real-space fractional coordinates, small geometric variations can lead to different serializations even for structures that share the same topological prototype. Figure~\ref{fig:2} illustrates this effect schematically. Structures with a common quotient graph may still receive different ESP orderings when small numerical differences alter the coordinate ranking. This introduces representation-level variability that is not directly related to the underlying topology.

ESG instead uses quotient-graph information to guide atomic ordering. To assess ordering variability across related structures, we grouped the MP-20 dataset using a strict structure-matching criterion and a coarse atom-count criterion, and counted the distinct ordered representations produced by each strategy (Table~\ref{tab:1}). In this test, smaller counts indicate lower serialization degeneracy. Here, the reported sequence variants are counted across all structures within each prototype cluster after chemistry homogenization, rather than across repeated perturbations of a single structure. These counts characterize variation across structures within each group after chemistry homogenization; the graph-construction counterexample in Supplementary Note S7 explains why geometric matching and ESG ordering can distinguish structures differently.

Overall, ESG yields fewer distinct sequence variants than ESP for most prototype clusters, indicating lower serialization degeneracy, although this advantage is not uniform in every case. For example, under strict clustering, prototype 6 produces 78 ESG variants compared with 264 under ESP, and prototype 2 yields 226 versus 486. Under loose clustering, prototype 10 yields 1,092 ESG variants versus 1,475. A few strict clusters show the reverse trend because geometric matching introduced by \emph{StructureMatcher} and Voronoi-derived graph equivalence distinguish structures differently (Supplementary Note S7). Across the evaluated structural families, ESG generally reduces ordering variability while retaining distinctions arising from the adopted connectivity definition.

\subsection{Unconditional Generation Benchmark}

After establishing that ESG reduces serialization variability across related structures, we next evaluated its impact on unconditional crystal generation. We first performed ablation studies to examine the roles and effects of the different ordering components (Supplementary Note S4 and Fig. S1). These results show that the element (E) and symmetry (S) operators provide useful chemical and crystallographic priors that improve compositional validity. Combining E and S with either P or G yields high generation validity, establishing the importance of chemical and crystallographic ordering; the impact of replacing coordinate ordering with graph-guided ordering is further assessed in the CSP and property-conditioned tasks below.

We then benchmarked the ESG-based CGTransformer on the MP-20 dataset against published representative generative baselines, including diffusion models, autoregressive language models and hybrid architectures (Table~\ref{tab:2}). Detailed benchmark settings for the unconditional generation experiments are provided in the Methods section.

On MP-20, CGTransformer with ESG achieves the highest reported overall validity in this comparison (92.76\%) while using only 86M parameters. It also outperforms the two ESP-based variants in overall validity. In addition, ESG maintains strong coverage (\(COV - R = 97.91\%\), \(COV - P = 99.95\%\)). These results indicate that topology-guided ordering improves validity without sacrificing sample diversity.

We further evaluated CGTransformer on the more challenging MPTS-52 dataset, which contains structures with up to 52 atoms per unit cell. Under ESG ordering, increasing model size led to improved performance, with the most notable gains observed in overall validity. This result suggests that additional model capacity remains useful for larger and more complex crystal systems. Considering the trade-off between performance and computational cost, we therefore used the 200M model in the subsequent MPTS-52 benchmarks.

Together, these results suggest that physically motivated sequence representations can substantially improve the efficiency of autoregressive unconditional crystal generation, especially under limited model capacity.

\subsection{Composition-conditioned Crystal Structure Prediction}

We next evaluated CGTransformer on the classical crystal structure prediction (CSP) task under compositional constraints. Given a chemical-formula prompt, the model generates \(N = 1\) or \(N = 20\) candidate structures, which are then compared with the ground-truth structure using match rate and root mean square error (RMSE). Table~\ref{tab:3} summarizes the results on the MP-20 and MPTS-52 datasets. To keep the main-text comparison consistent, we include only published end-to-end generative baselines, including representative diffusion models and autoregressive models. Detailed benchmark settings for the CSP experiments are provided in the Methods section.

On MP-20, our 86M-parameter CGTransformer equipped with ESG already shows competitive performance within a relatively lightweight GPT-2 framework. At the same 86M model size, replacing fractional-coordinate ESP ordering with ESG raises the single-shot match rate from 56.10\% to 61.92\%, exceeding the larger autoregressive baselines CrystaLLM 200M and Mat2Seq 200M. Under \(N = 20\), ESG attains a match rate of 75.40\%, again outperforming CrystaLLM and marginally surpassing Mat2Seq, showing that our graph-theory-guided ordering improves formula-conditioned structure recovery within the same autoregressive architecture.

To further examine the effect of model scaling on CSP, we additionally evaluated a 200M version of CGTransformer on MP-20. Under the same ESG ordering, scaling the model from 86M to 200M further improves performance, increasing the match rate from 61.92\% to 63.24\% for \(N = 1\), the highest among the published generative baselines listed in Table~\ref{tab:3}, and from 75.40\% to 76.43\% for \(N = 20\), while also reducing the RMSE from 0.0477 to 0.0438 and from 0.0532 to 0.0494, respectively. These results show that topology-aware representation and model capacity provide complementary gains: ESG establishes a better-aligned sequence space, while additional capacity further improves structure recovery within that space.

For the more challenging MPTS-52 benchmark, which contains substantially larger and more complex structures, we use the 200M CGTransformer as the main evaluation model. In this setting, ESG continues to outperform the diffusion baselines and remains competitive with previous autoregressive models. At the same 200M model size, ESG increases the \(N = 1\) match rate from 18.59\% to 21.78\% and reduces RMSE from 0.1226 to 0.0978 relative to fractional-coordinate ESP ordering. Its \(N = 1\) RMSE is the lowest among the main-text baselines, while its match rate remains close to Mat2Seq 200M. These comparisons demonstrate that sequence design improves CSP performance across both benchmark complexities without increasing model capacity.

In addition to structure-recovery accuracy, we quantified inference efficiency under the same MP-20 \(N = 20\ \) CSP tasks on a single NVIDIA A100 GPU (Table~\ref{tab:4}) as done in Mat2Seq\citep{ref29}. CGTransformer-small (86M) requires 2.0 s per crystal to generate 20 candidates while large version (200M) requires 2.5 s per crystal, which is substantially faster than CDVAE (37.9 s) and DiffCSP (7.3 s), slightly faster than Mat2Seq-small (2.1 s), and markedly faster than Mat2Seq-large (5.7 s). In this setting, CGTransformer-large attains an RMSE of 0.049, matching DiffCSP. These results indicate that CGTransformer provides a favourable inference speed for formula-conditioned crystal structure prediction.

\subsection{Property-guided inverse design}

We next evaluated CGTransformer-small (86M) in property-conditioned inverse design, focusing on the generation of wide-bandgap semiconductors under explicit bandgap constraints. As a proof-of-concept target, we consider materials with bandgaps in the range of 3.0--3.5 eV, which are relevant to applications such as solid-state lighting, power electronics and high-temperature devices. The model was first pretrained on the Alex-MP-20 dataset, which contains more than 600,000 structures, and was then fine-tuned on approximately 42,000 structures annotated with DFT bandgap values using Low-Rank Adaptation (LoRA, 2.4M trainable parameters). Additional fine-tuning details are provided in Supplementary Note S6.

The training data are strongly enriched in small-bandgap and metallic compounds (Supplementary Fig. S2). Despite this imbalance, bandgap-conditioned fine-tuning shifts the generated distribution towards the target interval. Under the 3.0--3.5 eV constraint, we generated 1,000 candidate structures using ESG and ESP, respectively. Following the evaluation protocol of MatterGen\citep{ref21}, the generated structures were relaxed and screened with MatterSim\citep{ref56} to obtain S.U.N. candidate structures and then validated using DFT calculations to assess electronic structure and bandgap values (Table~\ref{tab:5}).

ESG improves both screened candidate yield and bandgap targeting. Relative to ESP, it raises the S.U.N. rate from 12.9\% to 16.5\%, a relative increase of approximately 28\%; raises the target-window hit fraction among DFT-converged S.U.N. candidates from 17/120 (14.2\%) to 30/154 (19.5\%); and lowers the bandgap RMSE from 1.67 to 1.49. Here, hit rate is defined as the fraction of target-zone S.U.N. structures among all S.U.N. structures with converged DFT electronic-structure verification. The denominator is therefore smaller than the total number of S.U.N. candidates counted in the S.U.N. rate, because a small subset of S.U.N. candidates did not yield converged DOS/band-structure calculations after ten DFT retry attempts and was excluded from the hit-rate calculation (Supplementary Note S1). ESG also produces a smaller fraction of structures with bandgaps below 0.5 eV (18.2\% versus 21.7\%), suggesting reduced collapse towards the low-bandgap region that is enriched in the training data. The average energy above hull remains similar between the two strategies (0.096 versus 0.092 eV/atom), indicating that the improved bandgap targeting does not come with a large penalty in thermodynamic plausibility.

\begin{figure}[H]
\centering
\includegraphics[width=1\linewidth]{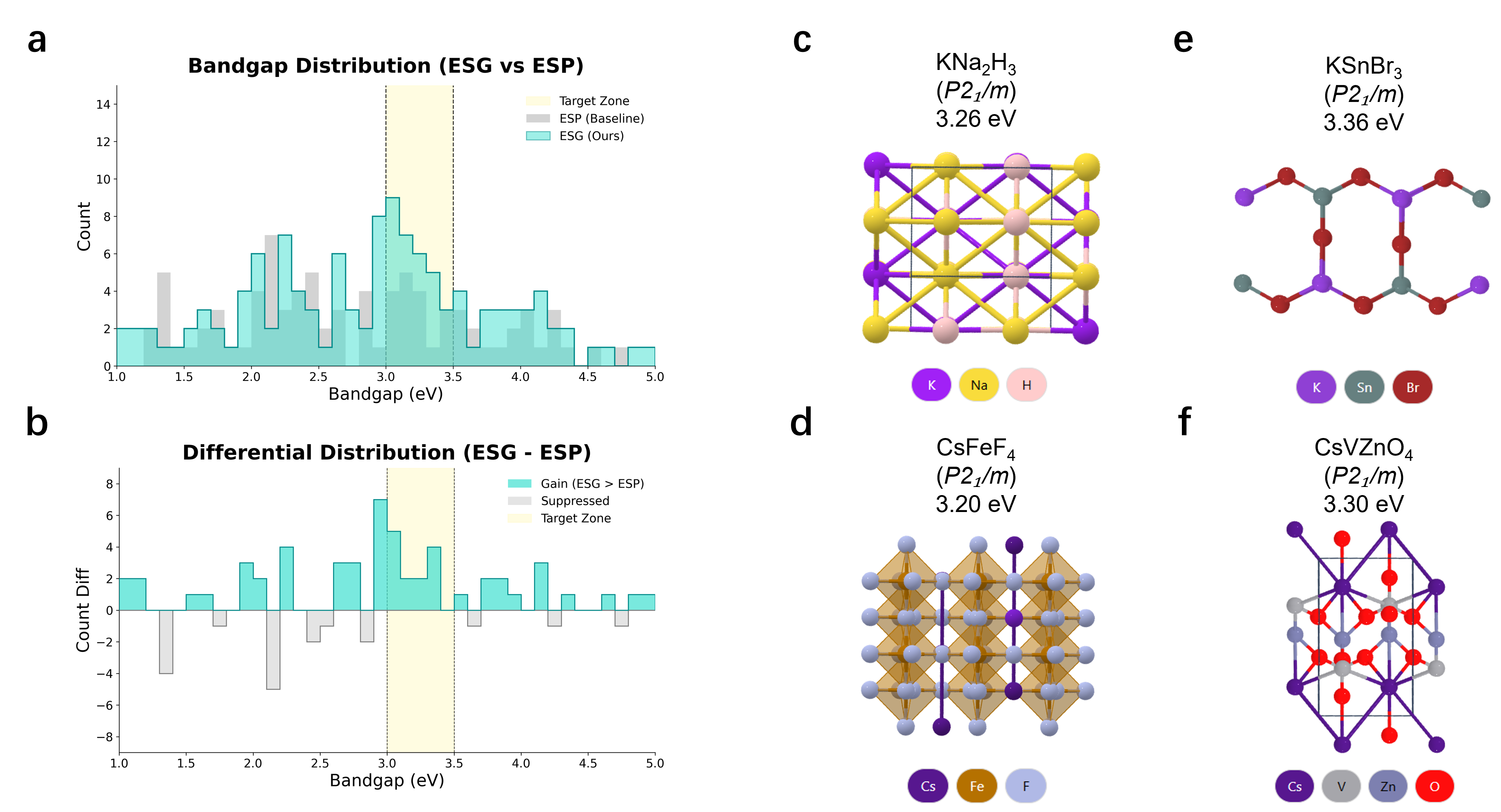}
\caption{\textbf{Property-conditioned inverse design of wide-bandgap semiconductors using CGTransformer.} \textbf{a,} DFT-computed bandgap distributions of the converged S.U.N. candidates generated using ESG (n=154) and ESP (n=120) under the 3.0--3.5 eV target condition. The displayed bandgap range is 1--5 eV. \textbf{b,} Difference between the corresponding bin counts (ESG minus ESP); positive values indicate bins in which ESG generates more structures. \textbf{c--f,} Representative S.U.N. structures generated by CGTransformer within the target bandgap interval, together with their key structural motifs and DFT-validated properties.}
\label{fig:3}
\end{figure}

Figure~\ref{fig:3}a,b illustrates the increase in screened candidates near the target interval. ESG produces more target-window candidates and achieves a higher target-window fraction, consistent with the increase from 17/120 to 30/154. The broader normalized distributions relative to the labelled training data are shown in Supplementary Fig. S2. Using the same 86M GPT-based architecture, topology-guided ordering improves alignment with the target bandgap interval at fixed model capacity, linking representation design directly to property-directed generation.

In Fig.~\ref{fig:3}c--f, we show four S.U.N. structures that fall within the target bandgap interval. These examples are chosen due to their novelty relative to the Materials Project (MP) database (until 2026): KNa\textsubscript{2}H\textsubscript{3} correspond to previously unreported stoichiometries within known chemical systems, KSnBr\textsubscript{3} \emph{P2\textsubscript{1}m} and CsFeF\textsubscript{4} \emph{P2\textsubscript{1}m} represents candidate polymorph not found among existing MP entries for that composition, and the Cs--V--Zn--O example lies outside the corresponding MP chemical system altogether. Notably, all four structures are assigned to the same space group under a Symmetry-finding tolerance set to 0.01 \AA{}, suggesting a non-random concentration of the generated candidates within a specific crystallographic family, even though this space group represents only a small fraction of MP structures in the relevant bandgap window (approximately 2\% of MP structures in the 3.2--3.4 eV bandgap window). This pattern suggests that ESG-guided generation is not merely producing isolated outliers, but may preferentially explore a structurally coherent region of wide-bandgap crystal space. A related concentration into specific structural families has also been reported in recent generative crystal-structure studies; for example, ternary chalcogenides generated by Chemeleon\citep{ref23} were found to be biased toward triclinic and monoclinic symmetries\citep{ref45}, consistent with the broader idea that crystal generative models may not sample structure space uniformly, but instead concentrate probability mass in chemically favorable regions.

\section{Discussion}

Our results show that topology-aware serialization substantially improves autoregressive modelling of crystalline materials. A central conclusion of this work is that sequence construction is not a neutral preprocessing step for crystal language modelling. Because periodic crystals do not possess a unique natural serialization, coordinate-based ordering can map closely related, or even isostructural, crystals to substantially different token sequences. This representation-level ambiguity weakens the learning signal by forcing the model to absorb sequence variability that is not directly tied to the underlying physics. ESG mitigates this effect by constructing a topology-aware sequence space that is more stable to local geometric perturbations and more aligned with local bonding connectivity.

More broadly, these findings suggest that strong crystal-generation performance can be achieved not only through architectural innovation, but also through careful representation design. When the input sequence is made more consistent with crystal topology, even a relatively standard GPT-based transformer can achieve strong performance across multiple tasks. These findings motivate the joint design of physical representations and autoregressive models. Periodic topology provides a concrete guide for organizing crystal sequences, while model scaling offers additional gains within the resulting representation. Crystal language models therefore benefit from treating sequence construction as part of model design, with measurable consequences for both structural recovery and property-directed generation.

One limitation of the current framework is that ESG depends on the adopted graph-construction rule. As illustrated by the counterexample discussed in Supplementary Note S7, structures that remain grouped together under geometric matching can still be distinguished by ESG if the inferred quotient-graph connectivity differs under the adopted Voronoi-based bonding convention. ESG is therefore deterministic under the specified graph-construction and starting-node protocol, and its stability is defined with respect to these choices.

\section{Methods}

\subsection{Dataset and Data Preprocessing}

To evaluate the generative performance of our model across multiple scales, we utilized three widely adopted standard crystallographic datasets: MP-20, MPTS-52, and Alex-MP-20. For unconditional generation and crystal structure prediction (CSP) tasks, we used a closed-set benchmarking protocol. The MP-20 dataset (originally introduced by CDVAE\citep{ref19}) was partitioned into training, validation, and test sets using a 60\%/20\%/20\% split ratio. The MPTS-52 dataset (introduced by DiffCSP\citep{ref20}, featuring more complex unit cells) was strictly divided into 27,380 / 5,000 / 8,096 structures for training, validation, and testing, respectively. For the CSP tasks, our models were trained from scratch exclusively on their respective benchmark training sets. For property-guided inverse design, we pretrained the model on the Alex-MP-20 dataset used in MatterGen\citep{ref21}. During the pre-training phase, the model was trained unconditionally on over 600,000 structures from this dataset. In the subsequent fine-tuning phase, a subset of approximately 42,000 structures annotated with density functional theory (DFT) bandgap labels was exclusively utilized for property-prompted conditional fine-tuning.

\subsection{CIF Syntax Standardization and Tokenization}

To bridge the representational gap between 3D physical structures and 1D natural language, we established a standardized "structure-to-sequence" transformation pipeline prior to model training. Initially, raw Crystallographic Information Files (CIFs) were processed using specific topological sorting algorithms (such as the proposed ESG) to ensure atom ordering canonicalization. The extracted core structural features---comprising three lattice lengths, three lattice angles, and the ordered atom types along with their corresponding 3D fractional coordinates---were restructured into JSON sequences. For conditional generation tasks, physical property labels (e.g., bandgap, space group, or bulk modulus) were prepended to these sequences as condition tokens. During the sequence discretization phase, we constructed a lightweight, custom-built tokenizer with a mixed-granularity vocabulary, including dedicated tokens for atom symbols, structural keywords, and special control markers, together with fine-grained tokens for numerical values and punctuation. The vocabulary of this tokenizer contains periodic table element symbols, numerical digits (rounded to one/two decimal places), property prefix identifiers, punctuation marks, and essential architectural special tokens. More detailed information could be found in Supplementary Note S2.

\subsection{Definition of Structure Prototypes}

To isolate the effect of the sorting strategy itself on sequence degeneracy, independent of chemical identity, we constructed two sets of structural prototype clusters from the MP-20 dataset. Specifically, all elements in each crystal were replaced with a single pseudo-atom type (for example, carbon), so that the analysis focused on structural variation rather than chemical composition.

To compare the sorting algorithms under a common initialization, dynamic permutation augmentation was disabled during this analysis. The first atom listed in the original CIF file was used as a common starting node, and each structure was translationally normalized by shifting this atom to the origin \((0,\ 0,\ 0)\), with all remaining atomic coordinates translated accordingly. This convention was introduced only to fix a shared initialization for comparing ESP and ESG.

The resulting structures were then grouped using two criteria.

\textbf{Strict structure-matching clustering.} Structural matching was performed using the StructureMatcher class in the pymatgen library. The tolerance parameters were set to standard values: fractional length tolerance (ltol) = 0.3, absolute site tolerance (stol) = 0.5 \AA{}, and angle tolerance (angle\_tol) = 10°. Following the clustering, the top 10 prototype groups containing the largest numbers of structures were selected to quantify the total number of unique sequence representations (Unique Sortings) generated by the ESP and ESG strategies, respectively.

\textbf{Loose size-based clustering.} A coarse-grained grouping was performed using only the number of atoms in the unit cell, covering all structures with \(N = 1\) to \(20\). For each size class, the numbers of distinct serialized sequences produced by the two ordering strategies were counted in the same manner.

\subsection{Benchmark Settings}

To comprehensively evaluate the generative performance of our model, we designed a three-stage evaluation pipeline encompassing unconditional generation, crystal structure prediction (CSP), and property-guided inverse design.

\textbf{Unconditional generation evaluation.} Utilizing model checkpoints trained for 500 epochs on the respective MP-20 and MPTS-52 training sets, we sampled 10,000 candidate structures for each dataset. Each generation process was initiated with a meaningless \textless bos\textgreater{} token. Following the settings established in CrystaLLM\citep{ref28} and Mat2Seq\citep{ref29}, the sampling strategy uniformly employed top-k sampling (k=30) with a temperature parameter set to 0.5 (t = 0.5). The evaluation framework primarily encompassed three dimensions: validity, coverage, and feature distance. First, physicochemical constraints were strictly enforced: structural validity mandates that the minimum distance between any pair of atoms must not be less than 0.5 \AA{}, effectively excluding unphysical atomic overlaps; compositional validity requires the generated chemical formula to satisfy charge neutrality based on standard oxidation states (evaluated via the SMACT algorithm\citep{ref57}). Second, to measure the similarity between the generated distribution and the ground-truth data distribution, we calculated Coverage Precision (COV-P) and Coverage Recall (COV-R) to quantify generation quality and diversity, respectively. We also report Wasserstein distances between generated and reference distributions of density (d\ensuremath{\rho}) and the number of distinct elements (d\textsubscript{elem}). Finally, we computed the Average Minimum Structure Distance (AMSD) and Average Minimum Composition Distance (AMCD) to further validate the distributional fidelity of the generated samples within a continuous, high-dimensional feature space. Complete metrics for unconditional tasks could be found in Supplementary Note S5 and Supplementary Table S3.

\textbf{Crystal structure prediction (CSP) evaluation.} For CSP, we evaluated 86M and 200M CGTransformer models on MP-20 and a 200M model on MPTS-52. During the inference phase, the models were conditioned exclusively on the chemical formulas from the test sets as prompts. For each given formula, the model was permitted to generate N=1 or 20 candidate structures. Consistent with the settings in CrystaLLM\citep{ref28} and Mat2Seq\citep{ref29}, a uniform top-k sampling strategy (k=10) was adopted with the temperature set to 1.0 (t = 1.0). To rigorously determine whether the generated structure matched the ground-truth structure and to calculate the root mean square error (RMSE), we employed the StructureMatcher algorithm from the pymatgen library\citep{ref58}. The matching tolerance parameters were strictly aligned with current domain standards: a fractional length tolerance (ltol) of 0.3, an absolute site tolerance (stol) of 0.5 \AA{}, and an angle tolerance (angle\_tol) of 10°. The average RMSE for the test set was computed by extracting the lowest RMSE value among all successfully matched candidates for each compound. Complete metrics for CSP tasks could be found in Supplementary Note S5 and Supplementary Table S4.

\textbf{Inverse design evaluation.} To assess the model\textquotesingle s property-conditioned inverse design capabilities, we set the challenge of generating wide-bandgap (3.0--3.5 eV) semiconductors. Following prompt fine-tuning, the model, guided by the corresponding property interval directives, randomly generated 1,000 crystal samples utilizing a top-k strategy (k=10) combined with a lower sampling temperature (t=0.8). Post-generation, we introduced a state-of-the-art machine learning universal force field (MatterSim, 5M parameters) to perform rapid structural relaxations and initial stability filtering on these candidate structures. Using MatterGen's released evaluation implementation, structures with a MatterSim-derived energy above the reference convex hull of \ensuremath{\leq}0.1 eV atom\textsuperscript{-1} were classified as stable. Uniqueness was assessed within the generated set, and novelty against the MP--Alexandria reference dataset with MP2020 energy corrections, using MatterGen's default ordered--disordered structure matcher. The stable, unique and novel (S.U.N.) screening rate was calculated as the fraction of all 1,000 generated structures satisfying all three criteria. Bandgaps of the screened S.U.N. candidates were then evaluated by static and non-self-consistent DFT electronic-structure calculations on the MatterSim-relaxed structures, using the PBE(+U) settings described in Supplementary Note S1.

\subsection{Sorting Methods and Algorithms}

\textbf{Sorting criteria:}

\textbf{E (Element-based):} Sorted by atomic species. Specifically, all atoms in the structure are ordered in ascending order of their atomic numbers.

\textbf{S (Symmetry-based):} Sorted by the symmetry of atomic positions. Specifically, the Wyckoff letters for all atoms in the structure are determined, and atoms are ordered alphabetically, which corresponds to an ascending order of position multiplicity.

\textbf{P (Position-based):} Sorted by the fractional coordinates of atoms. Specifically, the fractional coordinates of the starting atom (head atom) are set to (0, 0, 0), and the relative fractional coordinates of all other atoms are computed accordingly. If any relative coordinate falls outside the {[}0, 1) interval, it is mapped back to the {[}0, 1) range using periodic boundary conditions. Subsequently, the atoms are sorted in ascending order based on their \emph{x}-coordinates. If the \emph{x}-coordinates are identical, the \emph{y}-coordinates are compared to determine the order, followed by the \emph{z}-coordinates if necessary (i.e., a strict lexicographical order).

\textbf{G (Graph-based):} Sorted by the graph-embedded coordinates of the atoms. Specifically, inspired by previous works\citep{ref49,ref50,ref51,ref52}, O\textquotesingle Keeffe\textquotesingle s Voronoi method is used to determine the atomic bonding and construct the quotient graph of the structure. The coordinates of the starting node are set to (0, 0, 0), and the relative (fractional) coordinates of the remaining nodes are calculated using the standard barycentric embedding method. Taking the starting node as the root, a breadth-first search (BFS) algorithm traverses the graph, sorting nodes from shallowest to deepest based on their topological search depth. For nodes at the same depth, they are sorted in ascending order based on their standard embedded coordinates, similar to the logic in method P. Finally, the sorted nodes are mapped back to their corresponding atoms, defining the final atomic sequence.

\textbf{Combinatorial sorting strategies:}

\textbf{ESP (Element-Symmetry-Position):} In this case, order of different atoms will be determined by their atomic types (element), Wyckoff position (symmetry) and fractional/cartesian coordinates (fractional coordinates were used by default). First, method E is applied to group and sort atoms by element type. Next, method S is applied within each group to further sort atoms of the same species. At this stage, atoms are partitioned into subsets where both the atomic number and Wyckoff letter are strictly identical. Finally, method P is utilized to sort atoms within each subset, resolving spatial degeneracies and yielding the final sequence.

\textbf{ESG (Element-Symmetry-Graph):} Similar to the ESP strategy, methods E and S are sequentially applied to group and sort all atoms and then starting atom are determined. Finally, method G is applied to resolve remaining degeneracies within each subset, yielding the final sequence.

\textbf{Selection of the starting atom:} For a structure comprising N atoms, Head atom or starting atom would be chosen within the first atom subsets decided by method E and S, each individual atom could serve as the starting node, consequently generating distinct canonical sorting sequences.

\textbf{Construction of the labeled quotient graph via the VoronoiNN method}: In this work, we employ O\textquotesingle Keeffe\textquotesingle s method of Voronoi coordination (VoronoiNN)\citep{ref51} to determine the bonding relationships between atoms and to construct the labeled quotient graph. The VoronoiNN method utilizes a recommended\citep{ref52} default tolerance parameter, \(\delta = 0.5\), to adjust the number of nearest neighbors for each atom. For most structures, we use the default tolerance of 0.5 to construct the bonding network. However, for certain structures within the dataset, this default value fails to yield a fully connected labeled quotient graph, resulting in two or more disjoint subgraphs. For such cases, we iteratively decrease the tolerance parameter with a step size of 0.1 until a globally connected labeled quotient graph is successfully established.

\textbf{Barycentric embedding of periodic graphs}: The barycentric embedding of a periodic graph is a method for embedding graph nodes into Euclidean space, originally proposed by Delgado-Friedrichs and O\textquotesingle Keeffe\citep{ref49,ref50}. Following barycentric embedding, each node is geometrically positioned at the exact center of mass (barycenter) of its neighbors. Let \(p\) be the function that embeds the graph nodes \(\{ v\}\) into Euclidean space, and let \(v\) be a node in the graph. The embedded position \(p(v)\) satisfies the following relationship\citep{ref49,ref50}：
\[p(v) = \frac{1}{|N_{v}|}\sum_{w \in N_{v}} p(w)\]

Where \(N_{v}\) denotes the set of neighboring nodes of \(v\), and \(|N_{v}|\) represents the number of neighbors (i.e., the degree of node \(v\))

Given a labeled quotient graph, we can calculate its corresponding Laplacian matrix and the periodic offset-weighted degree matrix for all nodes. The coordinate matrix obtained after barycentric embedding satisfies the following linear system:
\[L \cdot C = O\]

For instance, considering the graphene-type structure. The Laplacian matrix of its relative quotient graph is formulated as:
\[L = \begin{pmatrix}
3 & - 2 & 0 & - 1 \\
 - 2 & 3 & - 1 & 0 \\
0 & - 1 & 3 & - 2 \\
 - 1 & 0 & - 2 & 3
\end{pmatrix}\]

For node 0, there are three incident edges connected to it. The weighted sum of its periodic offsets is precisely the vector sum of the periodic offsets of these three edges, namely (0, 0) + (0, 1) + (-1, 0) = (-1, 1). By calculating the weighted periodic offsets for all other nodes in a similar manner, we obtain the periodic offset-weighted degree matrix of the quotient graph:
\[O = \begin{pmatrix}
 - 1 & 1 \\
0 & - 1 \\
0 & 1 \\
1 & - 1
\end{pmatrix}\]

Let the barycentric embedded coordinates of node \(i\) in the quotient graph be \((x_{i},\ y_{i})\). By setting the coordinates of the first node to the origin \((0,\ 0)\), the relationship can be expressed according to the aforementioned equation as:
\[\begin{pmatrix}
3 & - 2 & 0 & - 1 \\
 - 2 & 3 & - 1 & 0 \\
0 & - 1 & 3 & - 2 \\
 - 1 & 0 & - 2 & 3
\end{pmatrix} \cdot \begin{pmatrix}
0 & 0 \\
x_{1} & y_{1} \\
x_{2} & y_{2} \\
x_{3} & y_{3}
\end{pmatrix} = \begin{pmatrix}
 - 1 & 1 \\
0 & - 1 \\
0 & 1 \\
1 & - 1
\end{pmatrix}\]

Solving this linear system yields the exact barycentric embedded coordinates for all nodes:
\[\begin{pmatrix}
x_{1} & y_{1} \\
x_{2} & y_{2} \\
x_{3} & y_{3}
\end{pmatrix} = \begin{pmatrix}
\frac{1}{6} & - \frac{1}{3} \\
\frac{1}{2} & 0 \\
\frac{2}{3} & - \frac{1}{3}
\end{pmatrix}\]

Because the atoms in the graphene structure inherently satisfy the conditions of barycentric embedding, the embedded coordinates computed in this example are mathematically equivalent to the true fractional coordinates of the atoms within the graphene crystal.

\clearpage

\section{Figures and tables}

\begin{table}[H]
\centering
\small
\setlength{\tabcolsep}{4pt}
\renewcommand{\arraystretch}{1.12}
\caption[Numbers of distinct serialized sequences within prototype clusters under ESG and ESP ordering.]{\textbf{Numbers of distinct serialized sequences within prototype clusters under ESG and ESP ordering.} Structures in the MP-20 dataset were first homogenized to a single pseudo-element and then grouped using either a strict structure-matching criterion (``Strict'') or a loose atom-count criterion (``Loose''). For each cluster, the number of distinct serialized sequences generated by ESG and ESP was counted under fixed initialization conditions. Smaller values indicate lower sequence degeneracy within the corresponding cluster. Bold values denote the smaller count in each row.}\label{tab:1}
\begin{tabular}{@{}lccclccc@{}}
\toprule
\multicolumn{4}{c}{\textbf{Strict}} & \multicolumn{4}{c}{\textbf{Loose}} \\
\textbf{Prototype} & \textbf{ESG} & \textbf{ESP} & \textbf{Num of Structures} & \textbf{Prototype} & \textbf{ESG} & \textbf{ESP} & \textbf{Num of Structures} \\
\midrule
1 & \textbf{6} & \textbf{6} & 2659 & 1 & \textbf{1} & \textbf{1} & 110 \\
2 & \textbf{226} & 486 & 854 & 2 & \textbf{1} & \textbf{1} & 961 \\
3 & 522 & \textbf{480} & 734 & 3 & \textbf{2} & \textbf{2} & 924 \\
4 & \textbf{6} & \textbf{6} & 591 & 4 & \textbf{6} & \textbf{6} & 6964 \\
5 & \textbf{6} & \textbf{6} & 573 & 5 & \textbf{24} & \textbf{24} & 2130 \\
6 & \textbf{78} & 264 & 514 & 6 & \textbf{117} & 120 & 3795 \\
7 & \textbf{23} & \textbf{23} & 513 & 7 & \textbf{181} & 255 & 991 \\
8 & \textbf{279} & 419 & 434 & 8 & \textbf{735} & 923 & 3563 \\
9 & 244 & \textbf{164} & 401 & 9 & \textbf{494} & 603 & 1527 \\
10 & \textbf{138} & 180 & 378 & 10 & \textbf{1092} & 1475 & 4395 \\
 &  &  &  & 11 & \textbf{305} & 309 & 590 \\
 &  &  &  & 12 & 1837 & \textbf{1815} & 4396 \\
 &  &  &  & 13 & \textbf{418} & 499 & 979 \\
 &  &  &  & 14 & \textbf{1347} & 1508 & 2959 \\
 &  &  &  & 15 & \textbf{334} & 346 & 666 \\
 &  &  &  & 16 & \textbf{1469} & 1520 & 3026 \\
 &  &  &  & 17 & \textbf{210} & 249 & 418 \\
 &  &  &  & 18 & \textbf{1184} & 1254 & 2418 \\
 &  &  &  & 19 & \textbf{242} & 268 & 468 \\
 &  &  &  & 20 & \textbf{2005} & 2020 & 3949 \\
\bottomrule
\end{tabular}
\end{table}

\begin{table}[H]
\centering
\small
\setlength{\tabcolsep}{3pt}
\renewcommand{\arraystretch}{1.12}
\caption[Results of unconditional generation benchmark]{\textbf{Results of unconditional generation benchmark.} CGTransformer with ESG ordering is compared with coordinate-based ESP variants and previously reported generative models. MP-20 results are compared against representative diffusion, autoregressive and hybrid baselines from the literature, whereas the MPTS-52 block reports internal comparisons under different ordering strategies and model scales. Higher values are better for validity and coverage metrics, and lower values are better for distance-based metrics. Bold values indicate the best value within each benchmark block.}\label{tab:2}
\begin{tabular}{@{}lccccccc@{}}
\toprule
 & \multicolumn{3}{c}{Validity(\%) \(\uparrow\)} & \multicolumn{2}{c}{Coverage(\%) \(\uparrow\)} & \multicolumn{2}{c}{Property \(\downarrow\)} \\
Method & Stru. & Comp. & overall & COV-R & COV-P & \(d_{\rho}\) & \(d_{elem}\) \\
\midrule
\multicolumn{8}{@{}l@{}}{\textbf{MP-20}} \\
\multicolumn{8}{@{}l@{}}{\textbf{Diffusion models}} \\
CDVAE & \textbf{100.0} & 86.70 & - & 99.15 & 99.49 & 0.6875 & 1.4320 \\
DiffCSP & \textbf{100.0} & 83.25 & - & 99.71 & 99.76 & 0.3502 & 0.3398 \\
UnitMat-Large & 97.20 & 89.40 & - & 99.80 & 99.70 & 0.0880 & 0.0560 \\
FlowMM & 96.85 & 83.19 & 80.30 & 99.49 & 99.58 & 0.239 & 0.083 \\
FlowLLM & 99.94 & 90.84 & 90.81 & 96.95 & 99.82 & 1.14 & 0.15 \\
\multicolumn{8}{@{}l@{}}{\textbf{Autoregressive models}} \\
CrystaLLM 25M & 94.97 & 93.80 & - & 97.58 & 95.97 & 1.1824 & 0.3269 \\
CrystaLLM 200M & 96.21 & \textbf{95.40} & - & 96.78 & 96.60 & 0.9835 & 0.3436 \\
Mat2Seq & 94.2 & 88.5 & - & - & - & - & - \\
\multicolumn{8}{@{}l@{}}{\textbf{Hybrid models}} \\
MP20-only ADiT & 99.58 & 90.46 & 90.13 & - & - & - & - \\
UniGenX 100 M & 99.08 & 90.12 & - & 99.27 & \textbf{99.95} & \textbf{0.065} & \textbf{0.040} \\
\multicolumn{8}{@{}l@{}}{\textbf{CGTransformer} (t=0.5)} \\
ESG (86M) & \ul{99.85} & \ul{92.91} & \textbf{92.76} & 97.91 & \textbf{99.95} & 0.9376 & 0.2675 \\
ESP (Fractional,86M) & 99.79 & 92.51 & 92.13 & 98.01 & 99.95 & \ul{0.7979} & 0.2456 \\
\shortstack{ESP\\(Cartesian, 86M)} & 99.81 & 92.30 & 91.31 & \ul{98.06} & 99.89 & 0.8325 & \ul{0.2182} \\
\multicolumn{8}{@{}l@{}}{\textbf{MPTS-52}} \\
\multicolumn{8}{@{}l@{}}{\textbf{CGTransformer} (t=0.5)} \\
Random & 89.42 & 76.22 & 69.61 & 99.46 & 99.56 & 0.4551 & 0.5112 \\
ESG (200M) & 95.01 & 86.69 & 82.22 & 99.42 & \ul{99.65} & \ul{0.4035} & \ul{0.4239} \\
ESG large (1.3B) & \ul{97.21} & \ul{89.16} & \ul{86.61} & \ul{99.48} & 99.37 & 0.6710 & 0.5933 \\
\bottomrule
\end{tabular}
\end{table}

\begin{table}[H]
\centering
\small
\setlength{\tabcolsep}{4pt}
\renewcommand{\arraystretch}{1.12}
\caption[Benchmark results for formula-conditioned crystal structure prediction on MP-20 and MPTS-52.]{\textbf{Benchmark results for formula-conditioned crystal structure prediction on MP-20 and MPTS-52.} Published diffusion and autoregressive generative models are compared with GPT-based models trained using different ordering strategies. For MP-20, CGTransformer is evaluated at both 86M and 200M parameters; for MPTS-52, a 200M model is used. For each composition prompt, models generate either one (\(N = 1\)) or twenty (\(N = 20\)) candidate structures. Performance is evaluated by match rate and RMSE relative to the ground-truth structure. Higher match rate and lower RMSE are better. Bold values denote the best result within each dataset block. Underlined values denote the best results of CGTransformer within each model scale.}\label{tab:3}
\begin{tabular}{@{}lcccc@{}}
\toprule
Methods & \multicolumn{2}{c}{N = 1} & \multicolumn{2}{c}{N = 20} \\
 & Match rate & RMSE & Match rate & RMSE \\
\midrule
\multicolumn{5}{@{}l@{}}{\textbf{MP20}} \\
\multicolumn{5}{@{}l@{}}{\textbf{Diffusion models}} \\
CDVAE & 33.90 & 0.1045 & 66.95 & 0.1026 \\
DiffCSP & 51.49 & 0.0631 & 77.93 & 0.0492 \\
FlowMM & 61.39 & 0.0566 & - & - \\
CrystalFlow & 62.02 & 0.0710 & \textbf{78.34} & 0.0577 \\
\multicolumn{5}{@{}l@{}}{\textbf{Autoregressive models}} \\
CrystaLLM 200M & 58.70 & 0.0408 & 73.97 & 0.0349 \\
Mat2Seq 200M & 61.30 & \textbf{0.040} & 75.30 & \textbf{0.037} \\
\multicolumn{5}{@{}l@{}}{\textbf{CGTransformer （86 M）}} \\
Random & 45.96 & 0.0620 & 69.05 & 0.0634 \\
ESP (Fractional) & 56.10 & 0.0525 & 72.64 & 0.0552 \\
ESP (Cartesian) & 53.47 & 0.0595 & - & - \\
ESG & \ul{61.92} & \ul{0.0477} & \ul{75.40} & \ul{0.0532} \\
\multicolumn{5}{@{}l@{}}{\textbf{CGTransformer （200M）}} \\
ESG & \textbf{63.24} & 0.0438 & 76.43 & 0.0494 \\
\multicolumn{5}{@{}l@{}}{\textbf{MPTS-52}} \\
\multicolumn{5}{@{}l@{}}{\textbf{Diffusion models}} \\
CDVAE & 5.34 & 0.2106 & 20.79 & 0.2085 \\
DiffCSP & 12.19 & 0.1786 & 34.02 & 0.1749 \\
FlowMM & 17.54 & 0.1726 & - & - \\
CrystalFlow & 21.00 & 0.1613 & 37.81 & 0.1584 \\
\multicolumn{5}{@{}l@{}}{\textbf{Autoregressive models}} \\
CrystaLLM 200M & 19.21 & 0.1110 & 33.75 & 0.1059 \\
Mat2Seq 200M & \textbf{23.1} & 0.109 & 36.4 & \textbf{0.088} \\
\multicolumn{5}{@{}l@{}}{\textbf{CGTransformer （200 M）}} \\
Random & 10.61 & 0.1155 & 22.38 & 0.1181 \\
ESP (Fractional) & 18.59 & 0.1226 & 31.53 & 0.1303 \\
ESG & \ul{21.78} & \textbf{\ul{0.0978}} & \ul{34.71} & \ul{0.1111} \\
\bottomrule
\end{tabular}
\end{table}

\begin{table}[H]
\centering
\small
\setlength{\tabcolsep}{4pt}
\renewcommand{\arraystretch}{1.12}
\caption[Efficiency and model complexity comparisons]{\textbf{Efficiency and model complexity comparisons.} Baseline timings are taken from Mat2Seq, all methods were evaluated by generating 20 candidate structures per composition on a single NVIDIA A100 GPU. RMSE is computed against the ground-truth target structure, and generation time is reported as the average wall-clock time per crystal. Lower RMSE and lower generation time indicate better performance.}\label{tab:4}
\begin{tabular}{@{}llcc@{}}
\toprule
MP-20 test set & Number of parameters & RMSE & \shortstack{20 shots generation\\speed (sec./crystal)} \\
\midrule
CDVAE & 4.5 M & 0.103 & 37.9 \\
DiffCSP & Similar to CDVAE & 0.049 & 7.3 \\
Mat2Seq-small & 25 M & 0.039 & 2.1 \\
Mat2Seq-large & 200 M & \textbf{0.037} & 5.7 \\
CGTransformer-small & 86 M & 0.053 & \textbf{2.0} \\
CGTransformer-large & 200 M & 0.049 & 2.5 \\
\bottomrule
\end{tabular}
\end{table}

\begin{table}[H]
\centering
\small
\setlength{\tabcolsep}{4pt}
\renewcommand{\arraystretch}{1.12}
\caption[Comparison of ESG and ESP for bandgap-conditioned inverse design of wide-bandgap semiconductors.]{\textbf{Comparison of ESG and ESP for bandgap-conditioned inverse design of wide-bandgap semiconductors.} Each method generated 1,000 candidate structures under a target bandgap interval of 3.0--3.5 eV. Structures were relaxed and screened using MatterSim. The mean MatterSim-derived energy above hull (eV atom\textsuperscript{-1}) and the stable, unique and novel (S.U.N.) screening rate are calculated over all 1,000 generated structures per method. All bandgap-based metrics---RMSE (eV), target-window hit fraction, and fractions with bandgaps below 0.5 eV or above 4.5 eV---are calculated from DFT results for converged S.U.N. candidates (154 for ESG and 120 for ESP). The target-window hit fraction is the proportion of these candidates with DFT bandgaps within 3.0--3.5 eV. Arrows indicate the preferred direction of each metric.}\label{tab:5}
\begin{tabular}{@{}lcc@{}}
\toprule
Method & ESG & ESP \\
\midrule
Ave. \(E_{hull}\) (eV/atom) \(\downarrow\) & 0.096 & \textbf{0.092} \\
S.U.N. rate \(\uparrow\) & \textbf{16.5}\% & 12.9\% \\
Hit rate \(\uparrow\) & \textbf{30/154} & 17/120 \\
RMSE \(\downarrow\) & \textbf{1.49} & 1.67 \\
Band gap \textless{} 0.5 eV \(\downarrow\) & \textbf{18.2 \%} & 21.7\% \\
Band gap \textgreater{} 4.5 eV \(\downarrow\) & 4.55\% & \textbf{3.33\%} \\
\bottomrule
\end{tabular}
\end{table}

\section{Code availability}

The code for training and using the CGTransformer is open source, released under the MIT License. The code repository is accessible online at Github (\url{https://github.com/Ky0Price/CGTransformer}) and Zenodo (10.5281/zenodo.20281899).

\section{Acknowledgments}

We especially thank Claudio Zeni, Senior Researcher at Microsoft Research Cambridge, AI for Science, for valuable discussions and advice throughout this study, drawing on his deep insights and extensive research experience. Calculations used supercomputers at the High-Performance Computing Center of the Collaborative Innovation Center of Advanced Microstructures and the High-Performance Supercomputing Center of Nanjing University.

\textbf{Funding:}

This work was supported by the National Natural Science Foundation of China (Grant Nos. 12125404, T2495231, and 12504277), the National Key R\&D Program of China (Grant No. 2022YFA1403201), the Advanced Materials-National Science and Technology Major Project (Grant No. 2024ZD0607000), the Basic Research Program of Jiangsu (Grant Nos. BK20233001, BK20241253, and BK20253009), the Fundamental and Interdisciplinary Disciplines Breakthrough Plan of the Ministry of Education of China (Grant No. JYB2025XDXM413), the Science Challenge Project (Grant No. TZ2025013), the AI \& AI for Science program of Nanjing University, the Artificial Intelligence and Quantum Physics (AIQ) program of Nanjing University, and the Fundamental Research Funds for the Central Universities. Y.H. acknowledges financial support from the Jiangsu Funding Program for Excellent Postdoctoral Talent (Grant No. 2025ZB852) and the Postdoctoral Fellowship Program of CPSF (Grant No. GZC20252202).

\textbf{Author contributions:}

Model architecture, training, and manuscript drafting: Z.Z. and H.Z. Algorithm development: S.Y. Code review and benchmark design: Y.H. DFT calculations: Z.G. Research conception and design, analysis, manuscript revision, and supervision: J.W., H.G., and J.S.

\textbf{Competing interests:}

The authors declare that they have no competing interests.

\textbf{Data availability:}

Additional data are available from the corresponding authors upon request by email.

\end{document}


\begin{center}
{\LARGE\bfseries \textbf{Supplementary Note for ``CGTransformer: A Topology-Aware Autoregressive Framework for Crystal Generation and Inverse Design''}\par}
\medskip
Zhennan Zhang\textsuperscript{1, \textdagger{}}, Shaobo Yu\textsuperscript{1, \textdagger{}}, Haoting Zhang\textsuperscript{1, \textdagger{}}, Yu Han\textsuperscript{1}, Zhiyue Guo\textsuperscript{1}, Junjie Wang\textsuperscript{1,*}, Hao Gao\textsuperscript{2,3,*}, Jian Sun\textsuperscript{1,*}\par
\smallskip
\small
\textsuperscript{1}~National Laboratory of Solid State Microstructures, School of Physics and Collaborative Innovation Center of Advanced Microstructures, Nanjing University, Nanjing 210093, China\par
\textsuperscript{2}~Guangdong Basic Research Center of Excellence for Structure and Fundamental Interactions of Matter, Guangdong Provincial Key Laboratory of Quantum Engineering and Quantum Materials, School of Physics, South China Normal University, Guangzhou 510006, China\par
\textsuperscript{3}~Guangdong-Hong Kong Joint Laboratory of Quantum Matter, Frontier Research Institute for Physics, South China Normal University, Guangzhou 510006, China\par
\end{center}
\begingroup
\renewcommand{\thefootnote}{}
\footnotetext[0]{\textsuperscript{\textdagger{}} These authors contributed equally to this work.

\textsuperscript{*}Corresponding author. J.S (\href{mailto:jiansun@nju.edu.cn}{\nolinkurl{jiansun@nju.edu.cn}}); H.G. (\href{mailto:gaohao@m.scnu.edu.cn}{\nolinkurl{gaohao@m.scnu.edu.cn}}); J.W. (\href{mailto:wangjunjie@nju.edu.cn}{\nolinkurl{wangjunjie@nju.edu.cn}})}
\endgroup

\tableofcontents

\clearpage

\section{DFT details}

All DFT calculations were performed using the Vienna ab initio simulation package (VASP)\citep{siref01} within the projector augmented-wave formalism via atomate2\citep{siref02}, pymatgen\citep{siref03} and custodian. All parameters of the calculations were chosen to be consistent with MP, including use of the PBE functional\citep{siref04} and Hubbard U corrections. Owing to the large computational cost of full structural optimization for the generated candidates, we did \textbf{not} perform DFT geometry relaxations in this verification stage. Instead, candidate structures were evaluated using static calculations followed by non-self-consistent band-structure calculations through the \emph{BandStructureMaker} workflow carried by atomate2. Band gaps reported in this work were extracted only from structures for which the static/non-self-consistent electronic-structure workflow completed successfully.

For the inverse-design evaluation, the ``10 attempts'' mentioned in the main text refer to \textbf{workflow-level resubmissions} of the static/non-self-consistent DFT electronic-structure verification. In each attempt, a candidate structure was processed through the same static + non-SCF band-structure workflow, with standard custodian-based error handling applied within that submission. Structures that still failed to complete this workflow after at most 10 attempts were classified as DFT-unverifiable and excluded from the hit-rate denominator.

\section{CIF Converting and Tokenization}

\subsection{CIF Converting}

The CIF format is highly flexible: the order of tags can vary arbitrarily, and not all tags are mandatory. In principle, a large language model could learn to process such files directly despite this variability. However, to reduce the learning burden imposed by heterogeneous file syntax, we converted all CIF files into a standardized JSON format and adopted a unified serialization and tokenization scheme for model input. In addition, auxiliary properties available in the dataset, such as dft\_band\_gap and energy\_above\_hull, were also included in the JSON files.

\textbf{Generation of JSON files}

The JSON files were constructed as follows:

\begin{enumerate}
\def\labelenumi{\arabic{enumi}.}
\item
  \textbf{Reading structural data.}\\
  All crystal structures used in this work were originally stored in CIF format. We first used the structure-reading function in ASE to parse each CIF file and convert it into an Atoms object.
\item
  \textbf{Sorting.}\\
  The parsed structure, containing \(N\) atoms, was then reordered using the corresponding sorting algorithm, yielding an ordered atomic sequence.
\item
  \textbf{Initializing the JSON record.}\\
  For each structure, the space-group identifier was first written into the JSON file.
\item
  \textbf{Writing the structural sequence.}\\
  For each ordered atomic sequence, the following information was written:
\end{enumerate}

\begin{quote}
\textbf{a. Element types and coordinates.}\\
The algorithm reads the atoms according to their indices in the ordered sequence and writes their element types together with their fractional coordinates into the JSON file. We further applied a simple normalization procedure to the coordinate data. Specifically, the coordinates of the first atom in the sequence were set to\((0,\ 0,\ 0)\), and the coordinates of all remaining atoms were expressed relative to this reference atom. For easier model learning, the relative coordinates were then mapped into the interval \(\left\lbrack 0,1 \right\rbrack\) and rounded to two decimal places.

\textbf{b. Lattice parameters.}\\
After the atomic sequence had been written, the six lattice parameters were recorded, including the three lattice lengths and the three lattice angles. For the same reason of facilitating model learning, all lattice parameters were rounded to one decimal place.
\end{quote}

An example of a CIF file from the training dataset, both before and after it has been pre-processed, is given below:

Original CIF file for mp-1185360 LiMnIr\textsubscript{2}, before pre-processing:

\begin{Verbatim}[fontsize=\small]
data_LiMnIr2
_symmetry_space_group_name_H-M   Fm-3m
_cell_length_a   5.88938400
_cell_length_b   5.88938400
_cell_length_c   5.88938400
_cell_angle_alpha   90.00000000
_cell_angle_beta   90.00000000
_cell_angle_gamma   90.00000000
_symmetry_Int_Tables_number   225
_chemical_formula_structural   LiMnIr2
_chemical_formula_sum   'Li4 Mn4 Ir8'
_cell_volume   204.27236420
_cell_formula_units_Z   4
loop_
 _symmetry_equiv_pos_site_id
 _symmetry_equiv_pos_as_xyz
  1  'x, y, z'
  2  '-x, -y, -z'
  3  '-y, x, z'
  ......
  192  '-x, z+1/2, -y+1/2'
loop_
 _atom_site_type_symbol
 _atom_site_label
 _atom_site_symmetry_multiplicity
 _atom_site_fract_x
 _atom_site_fract_y
 _atom_site_fract_z
 _atom_site_occupancy
  Li  Li0  4  0.00000000  0.00000000  0.50000000  1.0
  Mn  Mn1  4  0.00000000  0.00000000  0.00000000  1.0
  Ir  Ir2  8  0.25000000  0.25000000  0.25000000  1.0
\end{Verbatim}

Pre-processed JSON file for mp-1185360 LiMnIr\textsubscript{2}, only the first variant is shown below for brevity.:

\begin{Verbatim}[fontsize=\small]
{
        "dft_band_gap": 0.0,
        "energy_above_hull": 0.0180746750000002,
        "spacegroup": 225,
        ...... (other properties)
        "variants": [
            {
                "Ele0": "Li",
                "x0": 0.0,
                "y0": 0.0,
                "z0": 0.0,
                "Ele1": "Mn",
                "x1": 0.5,
                "y1": 0.5,
                "z1": 0.5,
                "Ele2": "Ir",
                "x2": 0.25,
                "y2": 0.25,
                "z2": 0.25,
                "Ele3": "Ir",
                "x3": 0.75,
                "y3": 0.75,
                "z3": 0.75,
                "a": 4.2,
                "b": 4.2,
                "c": 4.2,
                "alpha": 60.0,
                "beta": 60.0,
                "gamma": 60.0,
            },
(2 variants remain)
{
......
},
{
......
},
]
    }
\end{Verbatim}

\subsection{Tokenization}

After preprocessing, each structure stored in the JSON file is converted into a one-dimensional textual sequence. For example, a TiO\textsubscript{2} structure with \(Z = 2\) may be represented as following list:

{[}\textless bos\textgreater, unk\_prop, \ensuremath{\times}, 10, species, Ti, O, total\_atoms, 6, num\_atoms, 2, 4, \textless sep\textgreater, Ele0, Ti, x0, 0.0, y0, 0.0, z0, 0.0, Ele1, O, x1, ..., Ele5, O, x5, ..., a, ..., b, ..., c, ..., alpha, ..., beta, ..., gamma, ..., \textless eos\textgreater{]}

This sequence is tokenized into a sequence of integers, where each integer corresponds to a specific token. In our implementation, the tokenizer uses a \textbf{hybrid, mixed-granularity scheme} rather than a purely character-level representation. Frequently occurring structural fields, atom symbols, and control markers---such as species, total\_atoms, alpha, Ele, \textless bos\textgreater, \textless eos\textgreater, and unk\_prop---are represented as dedicated lexical tokens, while numerical values and punctuation marks are represented at a finer granularity. This design preserves the regular structure of crystal syntax while keeping the vocabulary compact and efficient for autoregressive modelling.

The vocabulary therefore contains three main categories of tokens::

\begin{itemize}
\item
  \textbf{numeric digits}
\item
  \textbf{alphabetic characters used in atomic symbols and structural tags}
\item
  \textbf{special symbols and punctuation}
\end{itemize}

More specifically, the vocabulary includes:

\begin{itemize}
\item
  \textbf{digits:} 0 1 2 3 4 5 6 7 8 9
\item
  \textbf{letters:} characters used in atomic symbols, property placeholders, structural tags, and space-group descriptors
\item
  \textbf{symbols and punctuation:} x y z . ( ) \textquotesingle{} , as well as whitespace and newline markers
\end{itemize}

The complete list of supported atomic species, structural tags, and space-group-related symbols in the token vocabulary is provided below as Table S1.

\begin{table}[H]
\centering
\small
\setlength{\tabcolsep}{4pt}
\renewcommand{\arraystretch}{1.12}
\caption[Token list]{\textbf{Token list}}\label{tab:s1}
\addcontentsline{toc}{subsection}{Table~\thetable. Token list}
\begin{tabular}{@{}cccccccccccc@{}}
\toprule
\multicolumn{12}{c}{\textbf{Full atom types (118)}} \\
\midrule
\textbf{H} & \textbf{He} & \multicolumn{2}{c}{\textbf{Li}} & \textbf{Be} & \textbf{B} & \textbf{C} & \textbf{N} & \multicolumn{2}{c}{\textbf{O}} & \textbf{F} & \textbf{Ne} \\
\textbf{Na} & \textbf{Mg} & \multicolumn{2}{c}{\textbf{Al}} & \textbf{Si} & \textbf{P} & \textbf{S} & \textbf{Cl} & \multicolumn{2}{c}{\textbf{Ar}} & \textbf{K} & \textbf{Ga} \\
\multicolumn{12}{c}{\textbf{\ldots\ldots{}}} \\
\multicolumn{12}{c}{\textbf{Special tokens}} \\
\multicolumn{3}{c}{\textbf{\textless pad\textgreater{}}} & \multicolumn{3}{c}{\textbf{\textless bos\textgreater{}}} & \multicolumn{3}{c}{\textbf{\textless eos\textgreater{}}} & \multicolumn{3}{c}{\textbf{\textless sep\textgreater{}}} \\
\multicolumn{12}{c}{\textbf{Key words}} \\
\multicolumn{2}{c}{\textbf{total\_atoms}} & \multicolumn{3}{c}{\textbf{num\_atoms}} & \multicolumn{2}{c}{\textbf{species}} & \multicolumn{3}{c}{\textbf{a}} & \multicolumn{2}{c}{\textbf{b}} \\
\multicolumn{2}{c}{\textbf{c}} & \multicolumn{3}{c}{\textbf{alpha}} & \multicolumn{2}{c}{\textbf{beta}} & \multicolumn{3}{c}{\textbf{gamma}} & \multicolumn{2}{c}{\textbf{Ele}} \\
\multicolumn{2}{c}{\textbf{unk\_prop}} & \multicolumn{3}{c}{\textbf{{[}}} & \multicolumn{2}{c}{\textbf{{]}}} & \multicolumn{3}{c}{\textbf{\{}} & \multicolumn{2}{c}{\textbf{\}}} \\
\multicolumn{2}{c}{\textbf{：}} & \multicolumn{3}{c}{\textbf{.}} & \multicolumn{5}{c}{\textbf{` '(blank space)}} & \multicolumn{2}{c}{} \\
\bottomrule
\end{tabular}
\end{table}

\section{Model Architecture and Generative Pre-training}

The generative pre-training step consists of training a GPT-style transformer model autoregressively. The implementation is based on the Hugging Face transformers library using the GPT-2 architecture. The model is composed of a stack of transformer blocks, each containing multi-head causal self-attention and a pointwise feed-forward network. The input to the model is a sequence of token indices derived from the tokenized crystal structure data.

Training consists of iteratively sampling sequences from the dataset and minimizing the cross-entropy loss between the predicted probability distribution and the actual next token. We utilize the AdamW optimizer with different learning rate values for different tasks. A linear learning rate scheduler is applied, which includes a warm-up phase of 500 steps followed by a linear decay over the course of training. The training utilizes an effective batch size of 32 (64 for alex-mp dataset) sequences per iteration. All models were trained on a CUDA-enabled GPU device with a fixed random seed of 2025 to ensure reproducibility.

\begin{table}[H]
\centering
\small
\setlength{\tabcolsep}{3pt}
\renewcommand{\arraystretch}{1.12}
\caption[Hyper parameters for different training tasks.]{\textbf{Hyper parameters for different training tasks.} The total number of parameters in parentheses is the number of trainable parameters}\label{tab:s2}
\addcontentsline{toc}{subsection}{Table~\thetable. Hyper parameters for different training tasks.}
\begin{tabular}{@{}lccccc@{}}
\toprule
Tasks & \shortstack{MP-20\\(86M)} & \shortstack{MPTS-52\\(200 M)} & \shortstack{MPTS52\\(1.34 B)} & Alex-MP20 Pretrain (86M) & \shortstack{Alex-MP20 Finetune\\(2.4 M)} \\
\midrule
Layer size & 12 & 16 & 44 & 12 & 12 \\
Attention heads & 12 & 16 & 25 & 12 & 12 \\
Embedding size & 768 & 1024 & 1600 & 768 & 1024 \\
Block size & 768 & 1024 & 1536 & 768 & 768 \\
Dropout & 0.1 & 0.1 & 0.1 & 0.1 & 0.1 \\
Batch size & 32 & 32 & 16 & 64 & 64 \\
\shortstack{Accumulate\\steps} & 1 & 1 & 2 & 1 & 1 \\
Learning rate & 5e-4 & 5e-4 & 3e-4 & 1e-3 & 1e-3 \\
Epochs & 500 & 500 & 500 & 50 & 50 \\
\bottomrule
\end{tabular}
\end{table}

\section{Ablation studies of different ordering components}

\begin{figure}[H]
\centering
\includegraphics[width=1\linewidth]{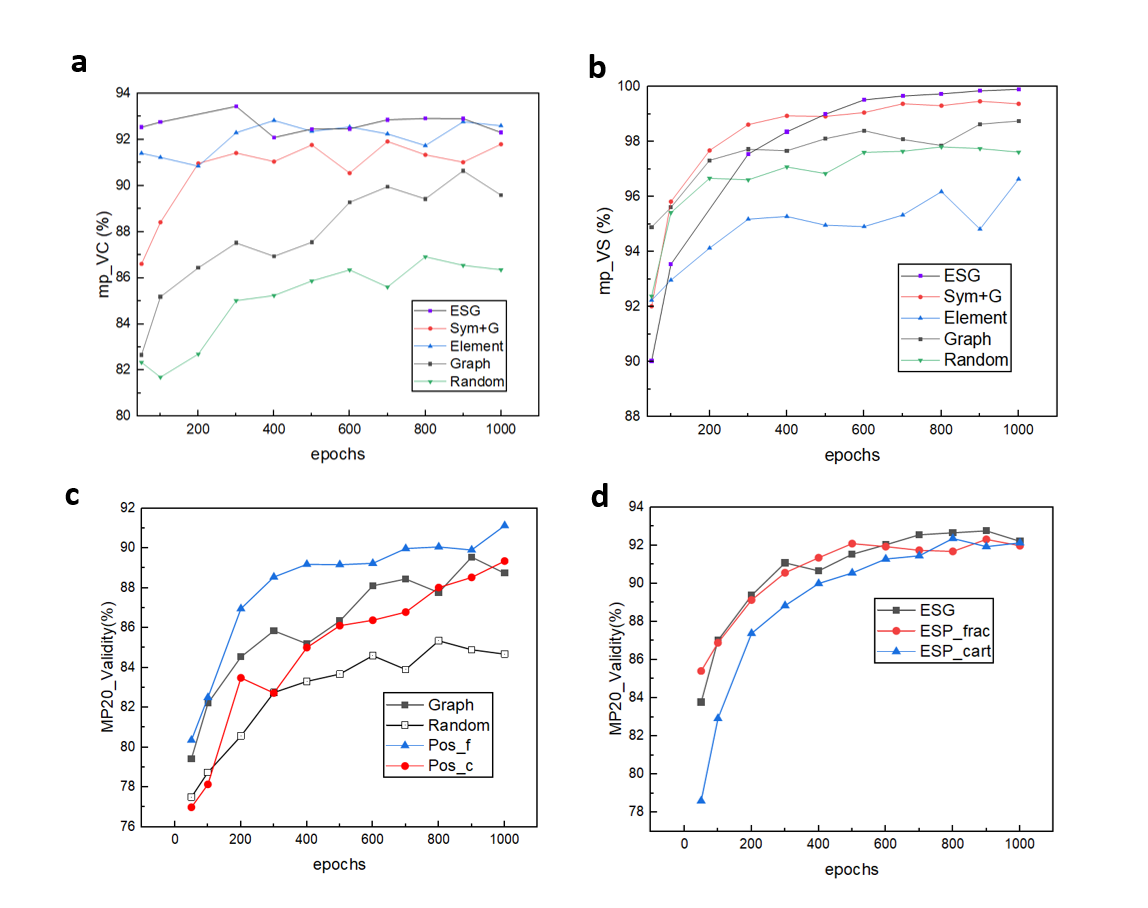}
\caption{\textbf{Validity--epoch curves for different ordering components and strategies in unconditional generation on MP-20.} \textbf{a.} Compositional validity versus training epoch for ESG, Sym+G, Element, Graph, and Random. \textbf{b.} Structural validity versus training epoch for the same ordering settings as in \textbf{a}. \textbf{c.} Overall validity versus training epoch for Graph, Random, Pos\_f, and Pos\_c. Here, Pos\_f and Pos\_c denote lexicographic sorting by fractional and Cartesian coordinates, respectively\textbf{. d.} Overall validity versus training epoch for the full combined strategies ESG, ESP\_frac, and ESP\_cart\textbf{.}}
\label{fig:s1}
\end{figure}

Fig~\ref{fig:s1} shows the validity--training-time curves of different ordering components and strategies for unconditional generation on the MP-20 dataset. As shown in Fig.~\ref{fig:s1}a,b, although the E (Element) and S (Symmetry) orderings are relatively simple in algorithmic design, both substantially improve the validity of the generated structures. In particular, the Element ordering contributes most strongly to composition validity, suggesting that a large part of the validity gain observed in the final ESG and ESP schemes can be attributed to the elemental prior introduced by E-ordering.

We further compared the standalone P and G orderings under unconditional generation (Fig.~\ref{fig:s1}c). Here, P-ordering refers to lexicographic sorting along \(x \rightarrow y \rightarrow z\), implemented either in fractional coordinates or in Cartesian coordinates, while the model input was kept in fractional coordinates for all settings. The results show that, when E and S are not incorporated, G-ordering alone does not exhibit a pronounced advantage. By contrast, P-ordering based on fractional coordinates performs noticeably better than Cartesian-coordinate sorting. One possible explanation is that some crystals exhibit highly anisotropic lattice vectors, in which case Cartesian-coordinate sorting becomes less stable than sorting in fractional space.

Finally, we compared the overall performance of ESG and ESP in the unconditional generation task (Fig.~\ref{fig:s1}d). Except in the early stage of training, where ESP-Cartesian performs slightly worse before approximately 500 epochs, the curves gradually converge as training proceeds. By the end of training, the differences between ESG and ESP in validity-related metrics become relatively small.

\section{Complete results of generation benchmarks}

To provide a more complete evaluation of unconditional crystal generation, we report not only validity metrics but also coverage- and distance-based metrics that quantify how well the generated distribution matches the reference dataset. Following the evaluation formalism introduced in previous crystal-generation benchmarks\citep{siref05}, we use six metrics to characterize different aspects of generation quality: \textbf{COV-R} (coverage recall), \textbf{COV-P} (coverage precision), \textbf{AMSD-R} (average minimum structure distance recall), \textbf{AMSD-P} (average minimum structure distance precision), \textbf{AMCD-R} (average minimum composition distance recall), and \textbf{AMCD-P} (average minimum composition distance precision).

Suppose a model generates a set of \(K\) crystal structures,
\[\{M_{k}\}_{k = 1}^{K} ,\]

and the corresponding ground-truth test set contains \(L\) crystal structures,
\[\{M_{l}^{*}\}_{l = 1}^{L}.\]

To compare these two sets, we define both a \textbf{structure distance} and a \textbf{composition distance} between pairs of crystals. Following CrystaLLM, the structure distance
\[D_{struc}(M_{k},M_{l}^{*})\]

is defined as the Euclidean distance between the CrystalNN-based structural fingerprints\citep{siref06} of the two structures, while the composition distance
\[D_{comp}(M_{k},M_{l}^{*})\]

is defined as the Euclidean distance between the normalized Magpie composition fingerprints\citep{siref07} of the two compounds. These two distances respectively measure similarity in local atomic environments and chemical composition.

Using dataset-dependent thresholds \(\delta_{struc}\) and \(\delta_{comp}\), the coverage metrics are defined as
\[\begin{aligned}
COV\text{-}R &= \frac{1}{L} \mid \left\{ l \in \lbrack 1,\ldots,L\rbrack:\exists k \in \lbrack 1,\ldots,K\rbrack,\text{ }\text{\,}D_{struc}(M_{k},M_{l}^{*}) < \delta_{struc},\; D_{comp}(M_{k},M_{l}^{*}) < \delta_{comp} \right\} \mid , \\
COV\text{-}P &= \frac{1}{K} \mid \left\{ k \in \lbrack 1,\ldots,K\rbrack:\exists l \in \lbrack 1,\ldots,L\rbrack,\text{ }\text{\,}D_{struc}(M_{k},M_{l}^{*}) < \delta_{struc},\; D_{comp}(M_{k},M_{l}^{*}) < \delta_{comp} \right\} \mid .
\end{aligned}\]

Here, \textbf{COV-R} measures how much of the reference data distribution is covered by the generated samples, whereas \textbf{COV-P} measures what fraction of generated samples falls sufficiently close to the reference distribution. Intuitively, recall emphasizes diversity and coverage of the ground-truth set, while precision emphasizes the quality and realism of generated samples.

The average minimum structure-distance metrics are defined as
\[\begin{aligned}
AMSD\text{-}R &= \frac{1}{L}\sum_{l = 1}^{L}\underset{k \in \lbrack 1,\ldots,K\rbrack}{\min}D_{struc}(M_{k},M_{l}^{*}), \\
AMSD\text{-}P &= \frac{1}{K}\sum_{k = 1}^{K}\underset{l \in \lbrack 1,\ldots,L\rbrack}{\min}D_{struc}(M_{k},M_{l}^{*}).
\end{aligned}\]

Similarly, the average minimum composition-distance metrics are defined as
\[\begin{aligned}
AMCD\text{-}R &= \frac{1}{L}\sum_{l = 1}^{L}\underset{k \in \lbrack 1,\ldots,K\rbrack}{\min}D_{comp}(M_{k},M_{l}^{*}), \\
AMCD\text{-}P &= \frac{1}{K}\sum_{k = 1}^{K}\underset{l \in \lbrack 1,\ldots,L\rbrack}{\min}D_{comp}(M_{k},M_{l}^{*}).
\end{aligned}\]

These four minimum-distance metrics provide a continuous assessment of distributional fidelity beyond hard threshold matching. AMSD-R and \textbf{AMCD-R} quantify how well the generated set covers the structural and compositional diversity of the reference set, whereas \textbf{AMSD-P} and \textbf{AMCD-P} quantify how close the generated samples are, on average, to the nearest reference structures in structure and composition space. Lower values of AMSD and AMCD indicate better agreement with the ground-truth distribution.

In summary, the recall-type metrics (\textbf{COV-R}, \textbf{AMSD-R}, \textbf{AMCD-R}) mainly evaluate whether the generated set spans the diversity of the true data distribution, whereas the precision-type metrics (\textbf{COV-P}, \textbf{AMSD-P}, \textbf{AMCD-P}) mainly evaluate whether the generated samples are concentrated in realistic regions of structure and composition space. Together with structural validity and compositional validity, these metrics provide a more complete picture of unconditional generation quality.

The threshold values \(\delta_{struc} = 0.4\) and \(\delta_{comp} = 10\) for MP-20 dataset used in this work follow the standard settings of the original benchmark protocol\citep{siref05}.

Table~\ref{tab:s3} presents the full set of metrics for unconditional generation, including the Average Minimum Distance family of metrics. These metrics complement the validity, coverage, and property-based statistics reported in the main text by providing a more stringent view of how closely the generated distribution matches the reference set.

Table~\ref{tab:s4} reports a more extensive comparison for the CSP task, including additional reference baselines, DiffCSP++\citep{siref08} and NatureLM\citep{siref09}. DiffCSP++ uses templates retrieved by CSPML\citep{siref10}, which retrieves templates using metric learning and then performs element substitution. NatureLM is trained on a broader corpus collected from Materials Project, NOMAD, and OQMD. We report these methods as supplementary references; the main-text comparison focuses on published end-to-end generative baselines.

\begin{table}[H]
\centering
\footnotesize
\setlength{\tabcolsep}{1.25pt}
\renewcommand{\arraystretch}{1.12}
\caption[Metrics for the unconditional generation tasks.]{\textbf{Metrics for the unconditional generation tasks.} In addition to the metrics reported in the main text, this table includes the \textbf{Average Minimum Distance} family of metrics, namely the \textbf{Average Minimum Structure Distance (AMSD)} and \textbf{Average Minimum Composition Distance (AMCD)}, which further evaluate the distributional fidelity of generated samples in a continuous high-dimensional feature space. The suffixes \textbf{-R} and \textbf{-P} denote \textbf{recall} and \textbf{precision}, respectively. Higher validity and coverage indicate better generation quality, whereas lower property and minimum-distance metrics indicate closer agreement with the reference distribution.}\label{tab:s3}
\addcontentsline{toc}{subsection}{Table~\thetable. Metrics for the unconditional generation tasks.}
\begin{tabular}{@{}lccccccccccc@{}}
\toprule
Method & \multicolumn{3}{c}{Validity(\%) \(\uparrow\)} & \multicolumn{2}{c}{Coverage(\%) \(\uparrow\)} & \multicolumn{2}{c}{Property \(\downarrow\)} & \multicolumn{4}{c}{Average Minimum Distance \(\downarrow\)} \\
 & Stru. & Comp. & overall & COV-R & COV-P & \(d_{\rho}\) & \(d_{elem}\) & AMSD-R & AMSD-P & AMCD-R & AMCD-P \\
\midrule
\multicolumn{12}{@{}l@{}}{\textbf{MP-20}} \\
\multicolumn{12}{@{}l@{}}{\textbf{Diffusion models}} \\
CDVAE & \textbf{100.0} & 86.70 & - & 99.15 & 99.49 & 0.6875 & 1.4320 & 0.154 & 0.188 & 3.620 & 4.014 \\
DiffCSP & \textbf{100.0} & 83.25 & - & 99.71 & 99.76 & 0.3502 & 0.3398 & - & - &  &  \\
UnitMat-Large & 97.20 & 89.40 & - & 99.80 & 99.70 & 0.0880 & 0.0560 & 0.097 & 0.119 & \textbf{2.410} & 2.410 \\
FlowMM & 96.85 & 83.19 & 80.30 & 99.49 & 99.58 & 0.239 & 0.083 & - & - & - & - \\
FlowLLM & 99.94 & 90.84 & 90.81 & 96.95 & 99.82 & 1.14 & 0.15 & - & - & - & - \\
\multicolumn{12}{@{}l@{}}{\textbf{Autoregressive models}} \\
CrystaLLM 25M & 94.97 & 93.80 & - & 97.58 & 95.97 & 1.1824 & 0.3269 & 0.106 & 0.095 & 3.729 & \textbf{1.762} \\
CrystaLLM 200M (t = 0.7) & 95.54 & 93.07 & - & 97.22 & 96.40 & 0.5965 & 0.1709 & \textbf{0.090} & 0.077 & 3.299 & 2.114 \\
CrystaLLM 200M (t = 0.5) & 96.21 & \textbf{95.40} & - & 96.78 & 96.60 & 0.9835 & 0.3436 & 0.098 & \textbf{0.076} & 3.675 & 1.880 \\
Mat2Seq & 94.2 & 88.5 & - & - & - & - & - & - & - & - & - \\
\multicolumn{12}{@{}l@{}}{\textbf{Hybrid models}} \\
MP20-only ADiT & 99.58 & 90.46 & 90.13 & - & - & - & - & - & - & - & - \\
UniGenX 100 M & 99.08 & 90.12 & - & 99.27 & \textbf{99.95} & \textbf{0.065} & \textbf{0.040} & - & - & - & - \\
\multicolumn{12}{@{}l@{}}{\textbf{CGTransformer} (86M, t=0.5)} \\
ESG & \ul{99.85} & \ul{92.91} & \textbf{92.76} & 97.91 & \textbf{99.95} & 0.9376 & 0.2675 & 0.097 & 0.080 & 3.597 & \ul{1.791} \\
ESP (Fractional) & 99.79 & 92.51 & 92.13 & 98.01 & 99.95 & \ul{0.7979} & 0.2456 & 0.097 & 0.082 & 3.614 & 1.812 \\
ESP (Cartesian) & 99.81 & 92.30 & 91.31 & \ul{98.06} & 99.89 & 0.8325 & \ul{0.2182} & \ul{0.094} & \ul{0.081} & \ul{3.526} & 1.831 \\
\multicolumn{12}{@{}l@{}}{\textbf{MPTS-52}} \\
\multicolumn{12}{@{}l@{}}{\textbf{CGTransformer} (t=0.5)} \\
Random (200M) & 89.42 & 76.22 & 69.61 & 99.46 & 99.56 & 0.4551 & 0.5112 & 0.174 & 0.148 & \ul{3.850} & 3.248 \\
ESP (200M) & 95.39 & 80.65 & 78.36 & 99.39 & \ul{99.65} & 0.4461 & 0.4649 & \ul{0.158} & \ul{0.135} & 3.889 & 3.110 \\
ESG (200M) & 95.01 & 86.69 & 82.22 & 99.42 & \ul{99.65} & \ul{0.4035} & \ul{0.4239} & \ul{0.158} & 0.136 & 3.937 & \ul{3.103} \\
ESG (1.34B) & \ul{97.21} & \ul{89.16} & \ul{86.61} & \ul{99.48} & 99.37 & 0.6710 & 0.5933 & 0.159 & 0.132 & 3.937 & 3.384 \\
Train(27380 crys) & 100.00 & 83.50 & 83.50 & - & - & - & - & - & - & - & - \\
\bottomrule
\end{tabular}
\end{table}

\begin{table}[H]
\centering
\small
\setlength{\tabcolsep}{4pt}
\renewcommand{\arraystretch}{1.12}
\caption[Metrics for the CSP tasks.]{\textbf{Metrics for the CSP tasks.} This table supplements the main-text comparison with additional reference baselines, including \textbf{DiffCSP++} , \textbf{NatureLM}.}\label{tab:s4}
\addcontentsline{toc}{subsection}{Table~\thetable. Metrics for the CSP tasks.}
\begin{tabular}{@{}lcccc@{}}
\toprule
 & \multicolumn{2}{c}{N = 1} & \multicolumn{2}{c}{N = 20} \\
 & Match rate & RMSE & Match rate & RMSE \\
\midrule
\multicolumn{5}{@{}l@{}}{\textbf{MP20}} \\
\multicolumn{5}{@{}l@{}}{\textbf{Diffusion models}} \\
CDVAE & 33.90 & 0.1045 & 66.95 & 0.1026 \\
DiffCSP & 51.49 & 0.0631 & 77.93 & 0.0492 \\
DiffCSP++ (w/ CSPML) & \textbf{70.58} & \textbf{0.0272} & - & - \\
FlowMM & 61.39 & 0.0566 & - & - \\
CrystalFlow & 62.02 & 0.0710 & \textbf{78.34} & 0.0577 \\
\multicolumn{5}{@{}l@{}}{\textbf{Autoregressive models}} \\
CrystaLLM 200M & 58.70 & 0.0408 & 73.97 & 0.0349 \\
Mat2Seq 200M & 61.30 & 0.040 & 75.30 & \textbf{0.037} \\
NatureLM-Mat3D 1B & 61.78 & 0.044 & - & - \\
\multicolumn{5}{@{}l@{}}{\textbf{CGTransformer （86 M）}} \\
Random & 45.96 & 0.0620 & 69.05 & 0.0634 \\
ESP (Fractional) & 56.10 & 0.0525 & 72.64 & 0.0552 \\
ESP (Cartesian) & 53.47 & 0.0595 & - & - \\
ESG & 61.92 & 0.0477 & 75.40 & 0.0532 \\
ESG (200 M) & \ul{63.24} & \ul{0.0438} & \ul{76.43} & \ul{0.0494} \\
\multicolumn{5}{@{}l@{}}{\textbf{MPTS-52}} \\
\multicolumn{5}{@{}l@{}}{\textbf{Diffusion models}} \\
CDVAE & 5.34 & 0.2106 & 20.79 & 0.2085 \\
DiffCSP & 12.19 & 0.1786 & 34.02 & 0.1749 \\
DiffCSP++ (w/ CSPML) & \textbf{37.17} & \textbf{0.0676} &  &  \\
FlowMM & 17.54 & 0.1726 & - & - \\
CrystalFlow & 21.00 & 0.1613 & 37.81 & 0.1584 \\
\multicolumn{5}{@{}l@{}}{\textbf{Autoregressive models}} \\
CrystaLLM 200M & 19.21 & 0.1110 & 33.75 & 0.1059 \\
Mat2Seq 200M & 23.1 & 0.109 & 36.4 & \textbf{0.088} \\
NatureLM-Mat3D 1B & 30.20 & 0.084 &  &  \\
\multicolumn{5}{@{}l@{}}{\textbf{CGTransformer（200M）}} \\
Random & 10.61 & 0.1155 & 22.38 & 0.1181 \\
ESP (Fractional) & 18.59 & 0.1226 & 31.53 & 0.1303 \\
ESG & \ul{21.78} & \ul{0.0978} & \ul{34.71} & \ul{0.1111} \\
\bottomrule
\end{tabular}
\end{table}

\section{Pretrain and finetune settings}

For pre-training, we follow the training pipeline of Mat2Seq. Specifically, we prepend ten placeholder property tokens (unk\_prop) to each CGTransformer crystal sequence and train the model on approximately 600,000 crystal structures. After pre-training, we fine-tune the model on a labelled dataset containing 42,000 crystal structures with corresponding bandgap values.

During fine-tuning, the bandgap values are discretized into 0.5 eV intervals and converted into condition tokens. For example, values in \(\lbrack 0,0.5)\) are assigned token 0, values in \(\lbrack 0.5,1.0)\) are assigned token 1, and so on. One unk\_prop token in the sequence is then replaced by the corresponding bandgap token. This design can be readily extended to other target properties or multiple targets by replacing the placeholder token with appropriate property tokens and fine-tuning the pretrained model.

For fine-tuning, we apply LoRA using the standard Hugging Face configuration for causal language models, with lora\_r = 16, lora\_alpha = 32, lora\_dropout = 0.05, and target\_modules = {[}"c\_attn", "c\_proj", "c\_fc"{]}. The remaining hyperparameters are summarized in table S2. When applied to the 86M model, this configuration results in 2.4M trainable parameters.

\begin{figure}[H]
\centering
\includegraphics[width=1\linewidth]{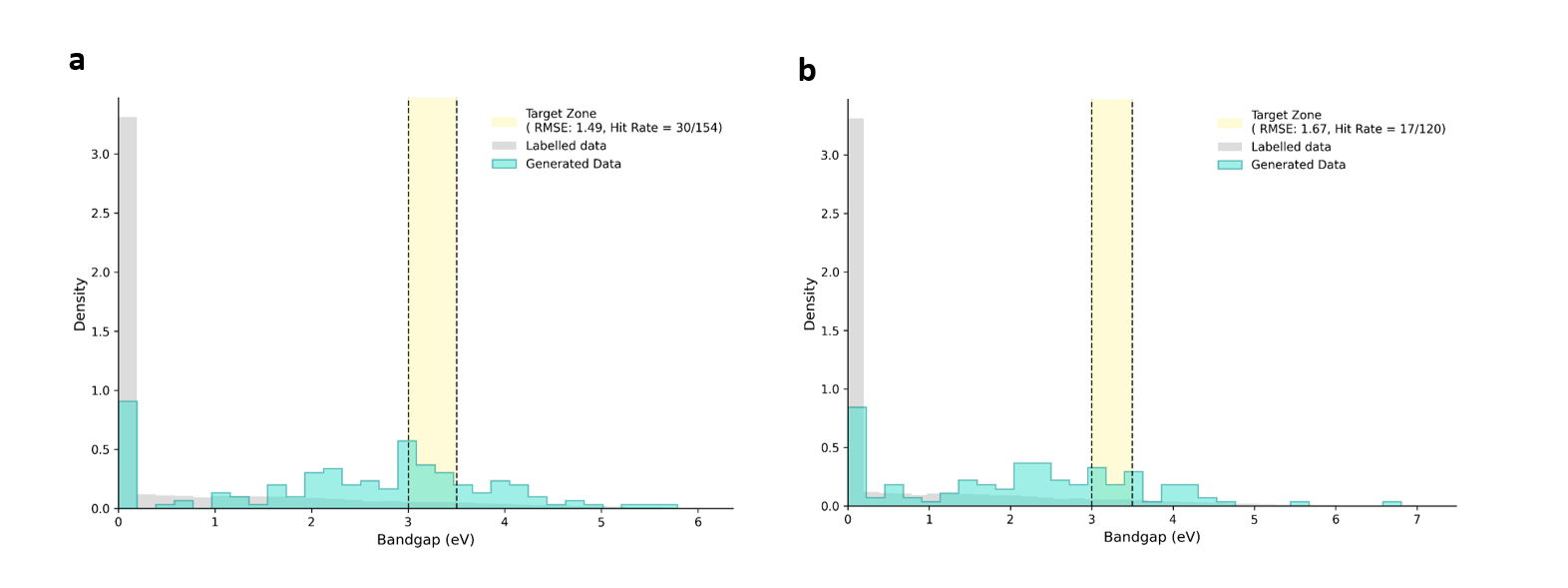}
\caption{\textbf{Distribution of generated data vs labelled data.} \textbf{a.} Distribution of ESG generated structures vs labelled data. \textbf{b.} Distribution of ESP generated structures vs labelled data}
\label{fig:s2}
\end{figure}

\section{A representative counterexample from the prototype analysis}

To clarify why ESG occasionally produces more sequence variants than ESP in a few strict prototype clusters, we analyze here a representative pair of structures from the strict prototype analysis. This pair was selected because it satisfies three conditions simultaneously: (i) the two structures are assigned to the same strict prototype cluster by \emph{StructureMatcher} under the default geometric tolerances, (ii) they yield the same ESP serialization, but (iii) they receive different ESG serializations.

\begin{figure}[H]
\centering
\includegraphics[width=0.88\linewidth]{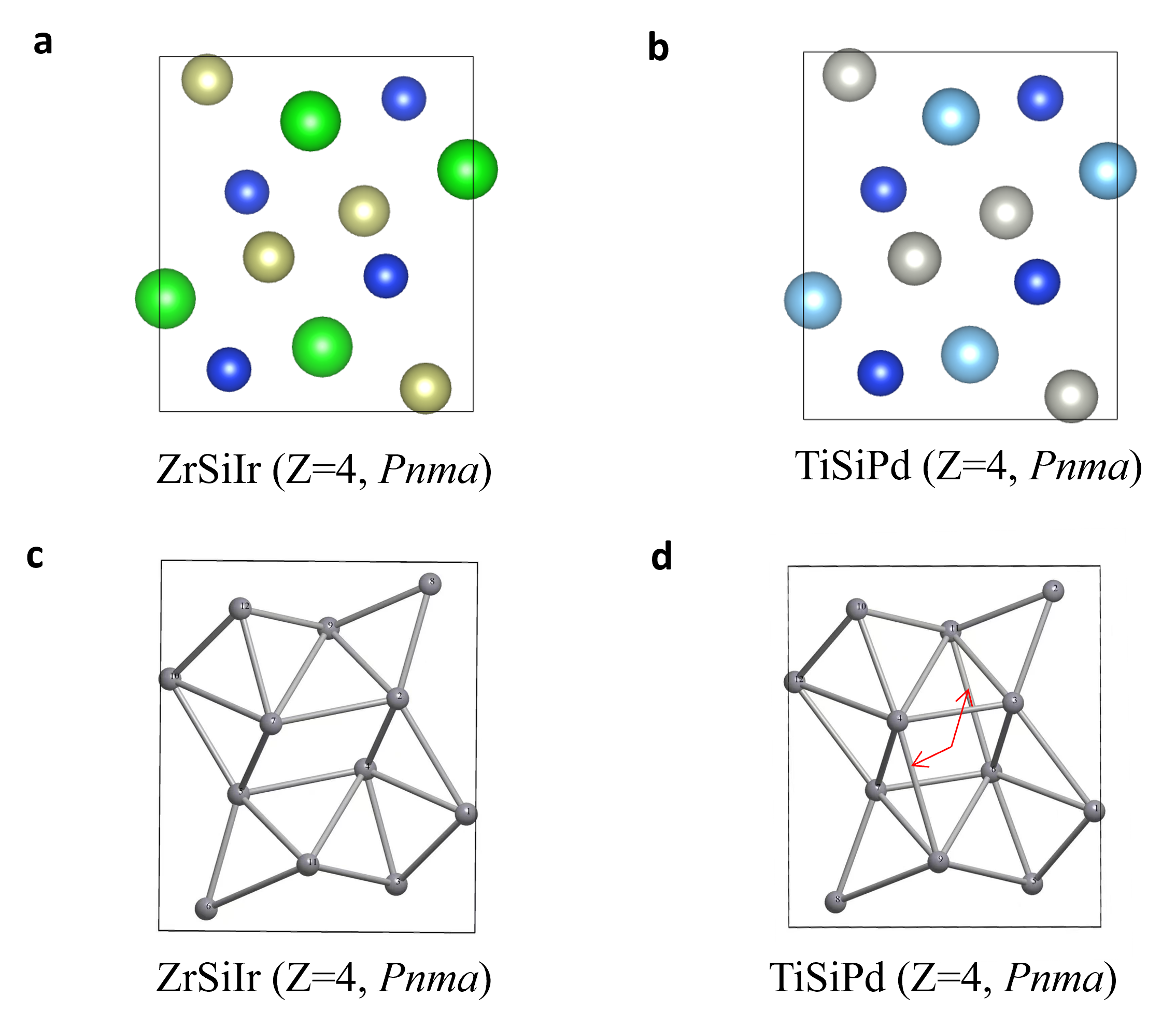}
\caption{\textbf{A representative counterexample from the prototype analysis.} \textbf{a.b.} Two structures assigned to the same strict prototype cluster by StructureMatcher under the default geometric tolerances and receiving the same ESP serialization. \textbf{c.d.} Connectivity inferred from the adopted O'Keeffe's VoronoiNN bonding criterion for the corresponding structures in \textbf{a} and \textbf{b}. Compared with the ZrSiIr structure, the TiSiPd structure contains two additional VoronoiNN-inferred graph edges (highlighted by red arrows), resulting in a different quotient graph and hence a different ESG serialization. This example illustrates that geometric clustering and graph-topological equivalence are closely related but not always identical}
\label{fig:s3}
\end{figure}

Figure~\ref{fig:s3} shows the corresponding structures and their inferred graph connectivity. The two crystals, ZrSiIr and TiSiPd, are geometrically similar enough to be grouped into the same prototype family by \emph{StructureMatcher} (Fig.~\ref{fig:s3}a,b). Under coordinate-based sorting, they also produce the same ESP ordering. However, after applying the adopted O'Keeffe's VoronoiNN bonding criterion to construct the quotient graph, the two structures no longer share identical connectivity. In particular, the TiSiPd structure contains two additional VoronoiNN-inferred graph edges relative to the ZrSiIr structure (Fig.~\ref{fig:s3}c,d), which leads to a different quotient graph and hence a different ESG serialization.

This example shows that the origin of these counterexamples is not the ordering stage itself, but the graph-construction stage used by ESG. In the prototype analysis, strict clustering is defined by geometric matching under StructureMatcher, whereas ESG relies on graph-topological equivalence under a separate bonding convention. These two notions of equivalence are related but not identical. As a result, structures that remain grouped together under geometric tolerances may still be distinguished by ESG if their inferred quotient-graph connectivity differs. These differences may be related to the tolerance parameter used in VoronoiNN neighbour selection, which can affect the inferred connectivity and hence the ESG ordering.

\phantomsection